\documentclass[12pt,prl,pas,showpacs,reprint]{revtex4-1}
\usepackage{graphicx}
\usepackage{geometry}
\usepackage{mathrsfs}
\usepackage{subfigure}
\usepackage{amssymb}
\usepackage{indentfirst}
\usepackage{color}
\usepackage{cancel}
\usepackage{amsmath}
\usepackage{hyperref}
\usepackage[]{appendix}
\usepackage[]{color}
\usepackage{float}
\begin{document}
\title{\LARGE\textbf{Optimized Large Hyperuniform Binary Colloidal Suspensions in Two Dimensions}}
\author{Zheng Ma}
\affiliation{Department of Physics, Princeton University, Princeton, New Jersey 08544, USA}
\author{Enrique Lomba}
\affiliation{Instituto de Qu\'{i}mica F\'{i}sica Rocasolano, CSIC,
Calle Serrano 119, E-28006 Madrid, Spain}
\author{Salvatore Torquato}
\affiliation{Department of Chemistry, 
Department of Physics, Princeton Institute for the Science and Technology of Materials,
and Program in Applied \\and Computational Mathematics, Princeton University,
Princeton, New Jersey 08544, USA}

\begin{abstract}
The creation of disordered hyperuniform materials with potentially extraordinary optical properties requires a capacity to synthesize large samples that are effectively hyperuniform down to the nanoscale.  Motivated by this challenge, we propose a fabrication protocol using binary paramagnetic colloidal particles confined in a 2D plane. The strong and long-ranged dipolar interaction induced by a tunable magnetic field is free from screening effects that attenuates long-ranged electrostatic interactions in charged colloidal systems. Specifically, we find a family of optimal size ratios that makes the two-phase system effectively hyperuniform. We show that hyperuniformity is a general consequence of low isothermal compressibilities, which makes our protocol suitable to systems with other long-ranged soft interactions, dimensionalities and/or polydispersity. Our methodology paves the way to synthesize large photonic hyperuniform materials that function in the visible to infrared range and hence may accelerate the discovery of novel photonic materials.
\end{abstract}
\date{}
\maketitle
{\it Introduction}.
Disordered hyperuniform systems are exotic disordered states that lie between crystals and liquids. While being statistically isotropic like liquids, they exhibit ``hidden order" in the sense that they suppress large-scale density fluctuations in the way that crystals do. Since the fundamental study on this subject \cite{torquato2003local}, hyperuniformity has been observed in a surprisingly wide variety of systems. Examples range from classical equilibrium systems \cite{dyson1962statistical, jancovici1981exact, torquato2015ensemble}, classical and quantum jammed systems \cite{PhysRevLett.95.090604, kurita2011incompressibility, zachary2011hyperuniformity, gerasimenko2019quantum}, critical absorbing states \cite{PhysRevLett.114.110602, weijs2015emergent, bertrand2019nonlinear}, active matter \cite{lei2019hydrodynamics, lei2019nonequilibrium}, soft polymers \cite{chremos2018hidden}, biological systems \cite{PhysRevE.89.022721, mayer2015well} to the one-dimensional point patterns derived from the nontrivial zeros of the Riemann zeta function \cite{montgomery1973pair}; see also the recent review for a more comprehensive list \cite{torquato2018hyperuniform}. Hyperuniform point configurations in $d$-dimensional space $\mathbb{R}^d$ possess a structure factor $S(\mathbf k)$ that goes to zero as the wavenumber $|\mathbf k|$ vanishes, i.e., $\lim_{|\mathbf k|\rightarrow 0} S(\mathbf k)=0$, which corresponds to a local number variance $\sigma_N^2(R)$ in a spherical window of radius $R$ that grows slower than $R^d$. The hyperuniformity concept has been generalized to two-phase media \cite{PhysRevE.94.022122}, where hyperuniformity means that the spectral density $\tilde\chi_{_V}(\mathbf k)$ (Fourier transform of the autocovariance function \cite{torquato2013random}) goes to zero as $|\bf k| \to 0$, which is equivalent to a local volume-fraction variance $\sigma_{_V}^2(R)$ that decreases faster than $R^{-d}$ for large $R$.

Besides being of great fundamental interest, hyperuniform materials are showing exciting technological promise, especially in photonics \cite{florescu2009designer, man2013isotropic, PhysRevA.88.043822, C4CP06024E, piechulla2018fabrication, bigourdan2019enhanced, gorsky2019engineered}. Specifically, disordered hyperuniform structures are found to exhibit photonic band gaps, just like photonic crystals, but possess the advantage of being isotropic, enabling free-form waveguides \cite{man2013isotropic}. These investigations motivated the study of hyperuniform materials for acoustic applications as well \cite{romero2019stealth}. Moreover, hyperuniform materials may be potentially useful for producing vivid non-iredescent structural colors \cite{noh2010noniridescent, chung2012flexible}.

However, a fundamental challenge is how to make large hyperuniform samples efficiently, especially down to the nanoscale. Computational protocols, such as the extensively used collective-coordinate optimization technique \cite{torquato2015ensemble}, involve high computational costs, which makes it hard to generate sample sizes beyond thousands of particles. Thus, a bottom-up, self-assembly based fabrication method is highly desired. Notable examples include jamming of hard or soft particles \cite{PhysRevLett.95.090604, ricouvier2017optimizing, kurita2011incompressibility, berthier2011suppressed}, periodically driven systems going through an absorbing phase transition \cite{weijs2015emergent, weijs2017mixing, wilken2020hyperuniform} and spinodal decompositions \cite{ma2017random,salvalaglio2019hyperuniform}.  The fact that many of these hyperuniform nonequilibrium systems are at critical points implies that there is little room to tune the structure, e.g., it is impossible to demand a significantly lower volume fraction in jammed systems. Moreover, any imperfections or defects, which always occurs in experiments, can degrade hyperuniformity, e.g., the rattlers in jammed packings \cite{atkinson2016critical}. 

On the other hand, equilibrium systems are much more robust and flexible in the sense that their macroscopic properties are time-independent and can be tuned by many parameters. Specifically, in order to observe hyperuniformity in an equilibrium system, the isothermal compressibility relation $S(0^{+})=\rho k_B T\kappa_T$ dictates a vanishing compressibility $\kappa_T$ at finite temperature $T$. This relation implies that a long-ranged interaction must be present. While systems that utilize Coulombic interactions may at first glance appear to be suitable to achieve hyperuniformity
(e.g., charged colloids), they often suffer from screening effects that attenuate the associated long-range electrostatic interactions. It has been shown that such systems only become effectively hyperuniform \cite{chen2018binary} at low temperatures and small inverse screening lengths.

In this Letter, we propose a highly feasible and robust equilibrium protocol that can be used in the laboratory to fabricate large disordered hyperuniform materials down to the nanoscale. Specifically, we consider superparamagnetic colloidal particles (doped with magnetic materials) confined at a two-dimensional interface such that dipole-dipole interactions are induced when a magnetic field is applied perpendicular to the plane. The interactions in such systems are strong, long-ranged ($u(r)\sim 1/r^3$) and free of screening effects, making them excellent candidate systems that can yield large, effectively disordered hyperuniform colloidal systems at positive temperatures. We actually employ binary particle mixtures to frustrate crystallization \cite{ebert2009partial}. The monodisperse version of our model has been extensively studied both numerically and experimentally \cite{lin2006computer, kapfer2015two, zahn1999two, kelleher2017phase}, mainly for the purpose of probing the nature of two-dimensional melting \cite{halperin1978theory}. Much less is known about the polydisperse case. Here we apply both Monte Carlo simulations and the Ornstein-Zernike integral-equation formalism to study the structure of the system. Similar techniques were applied in Ref. \cite{Hoffman2006}, but hyperuniformity was not a consideration. Moreover, here we focus on the two-phase systems formed by the particles, which is crucial for our purposes, as detailed below.

Our main finding is that in the equilibrium liquid phase, despite the structure of each component not being hyperuniform, there exists an optimal size ratio $R_1/R_2=(\rho_2S_{22}(0^{+})/\rho_1S_{11}(0^{+}))^{\frac{1}{4}}$ that makes the resulting two-phase system effectively hyperuniform. Here $S_{ii}$ and $\rho_i$ are the partial structure factor and density of the species $i$. The optimal size ratio leverages the destructive interference between scattering events between the two species, which is a general direct consequence of low isothermal compressibilities due to the strong and long-ranged repulsion. Our protocol can be potentially applied to other long-ranged soft repulsions ($u(r)\sim 1/r^n$), as they are shown to bear similar physics \cite{kapfer2015two}. Additionally, our protocol is also suitable to systems with other dimensionalities and/or polydispersity.

{\it Methods}. We study binary superparamagnetic colloidal particles confined in a 2D plane \cite{Hoffman2006, hoffmann2006partial, ebert2009partial}. An external magnetic field $\mathbf B$
perpendicular to the plane is applied, inducing a dipole-dipole interaction. For the binary system we consider here, denote $N_i$ and $\chi_i$ as the corresponding particle number and susceptibility for each component,
$i=1, 2$, with $\rho_i$ and $x_i$ being the number density and
concentration. We denote the species 1 as the ``large" particles and 2
the ``small" particles, in the sense that $\chi_1>\chi_2$. The dipole-dipole interaction $u_{ij}$ can be rewritten as  
$
\beta u_{ij}(x)={\Gamma_{ij}}/{x^3},
$
where $\beta=1/(k_BT)$, with $k_B$ being Boltzmann's constant, and $x$ is the distance between two particles rescaled by the
average interparticle distance between large particles, i.e., $x\equiv
r/a_{11}$, where $a_{11}=1/\sqrt{\rho_1}$. The quantity $\Gamma_{ij}$
is a dimensionless coupling strength between species $i$
and $j$, which can be written as 
$
\Gamma_{ij}=\beta{\mu_0\chi_i \chi_jB^2}/(8\pi a_{11}^3).
$
The binary colloids have radii $R_1$ and $R_2$ respectively, however in our simulations they can be treated conveniently as point-like due to the strong dipole-dipole repulsion \cite{hoffmann2006partial}. 
We perform Monte Carlo simulations of a system consists of 3600 particles, using a simplified swap Monte Carlo algorithm \cite{grigera2001fast}. Up to $10^9$ steps were used to equilibrate the systems, and all results presented are averaged using 30 to 50 configurations. We also solve the Ornstein-Zernike integral equation numerically to obtain the partial structure factors using the Rogers-Young \cite{Rogers1984} (RY) closure, see Ref.~\cite{Lomba2017} and Supplemental Material \cite{supplement} for details of the corresponding algorithm. 

{\it Results}.         
Importantly, we first show the binary system is not hyperuniform as a point pattern, but there exists a unique way of decorating the points to be effectively hyperuniform as a two-phase system. As noted above, hyperuniformity, i.e., the vanishing of $S(0^+)$ of an equilibrium one-component system is directly related to its incompressibility. However, we need to consider the isothermal compressibility $\kappa_T$ of a binary system, which can be expressed in terms of the
partial structure factors $S_{ij}(k)$ \cite{ashcroft1967structure}: 
\begin{equation} \label{compress}
\rho k_BT\kappa_T=\frac{S_{11}(0^{+})S_{22}(0^{+})-S_{12}^2(0^{+})}{x_1S_{22}(0^{+})+x_2S_{11}(0^{+})-2x_1^{\frac{1}{2}}x_2^{\frac{1}{2}}S_{12}(0^{+})},
\end{equation} 
where $S_{ij}(k)=(N_iN_j)^{-\frac{1}{2}}\left \langle \tilde n_i(k) \tilde n_j^*(k)\right \rangle$. Here $\tilde n_1(\mathbf k)$ and $\tilde n_2(\mathbf k)$ are complex collective density variables for large and small particles, which are defined as
$\tilde n_1(\mathbf k)\equiv \sum_{i=1}^{N_1}e^{-i\mathbf k \cdot \mathbf r_1{_i}}$
and
$\tilde n_2(\mathbf k)\equiv \sum_{i=1}^{N_2}e^{-i\mathbf k \cdot \mathbf r_2{_i}},$
where $\{r_1{_i}\}$ and $\{r_2{_i}\}$ refer to the set of positions of large and small particles. Incompressibility ($\kappa_T=0$) at positive
temperature then implies that
\begin{equation} \label{S_12}
S_{11}(0^{+})S_{22}(0^{+})-S_{12}^2(0^{+})=0.
\end{equation}
While incompressibility cannot be perfectly achieved in the dipolar system we study, due to the strong and long-ranged repulsion, this condition will approximately hold when the coupling strength $\Gamma$ is sufficiently large. 

\begin{figure}[h]
\centering
\subfigure[\ $\Gamma_{22}=1$]{
\includegraphics[width=4cm, height=3cm]{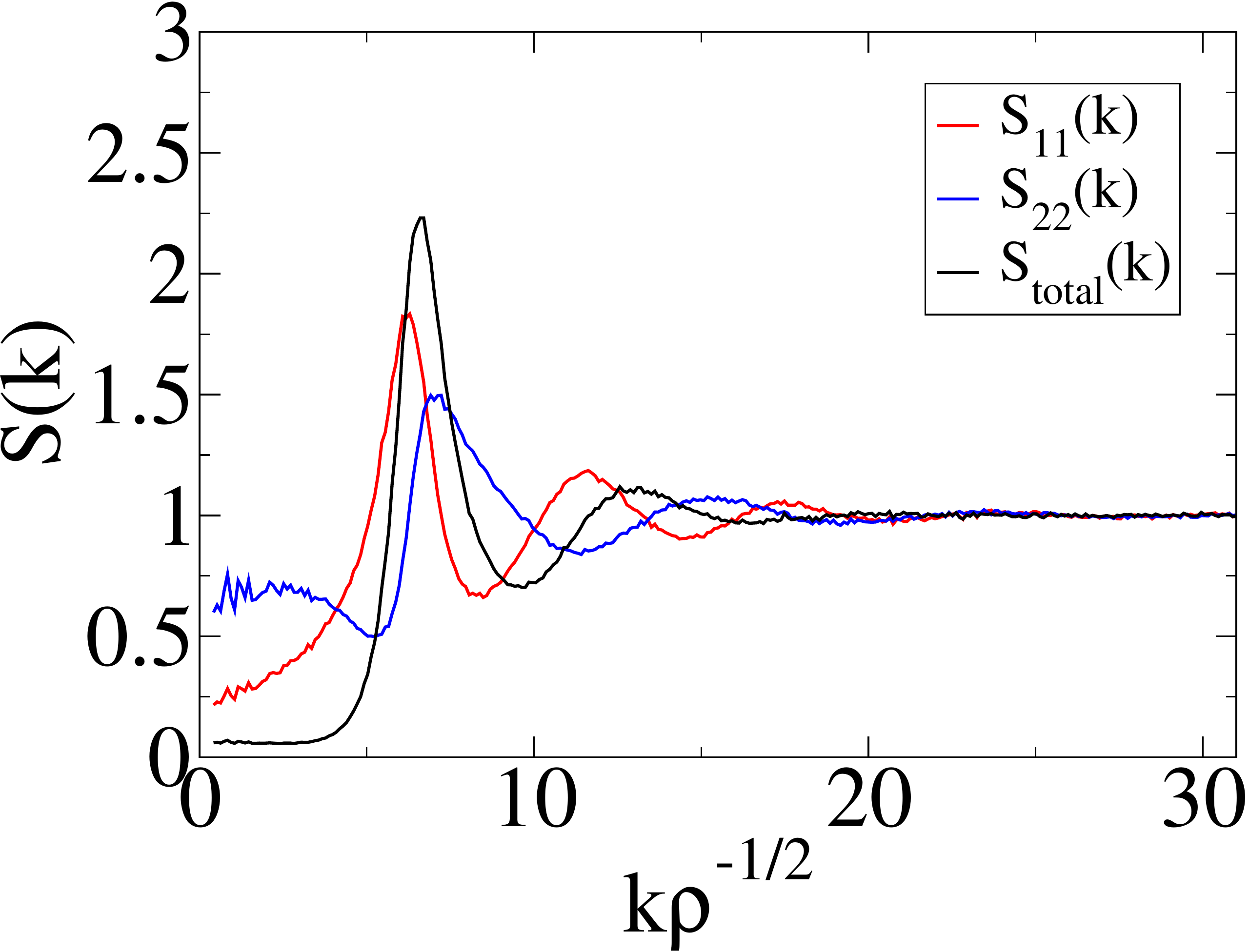}
}
\subfigure[\ $\Gamma_{22}=3$]{
\includegraphics[width=4cm, height=3cm]{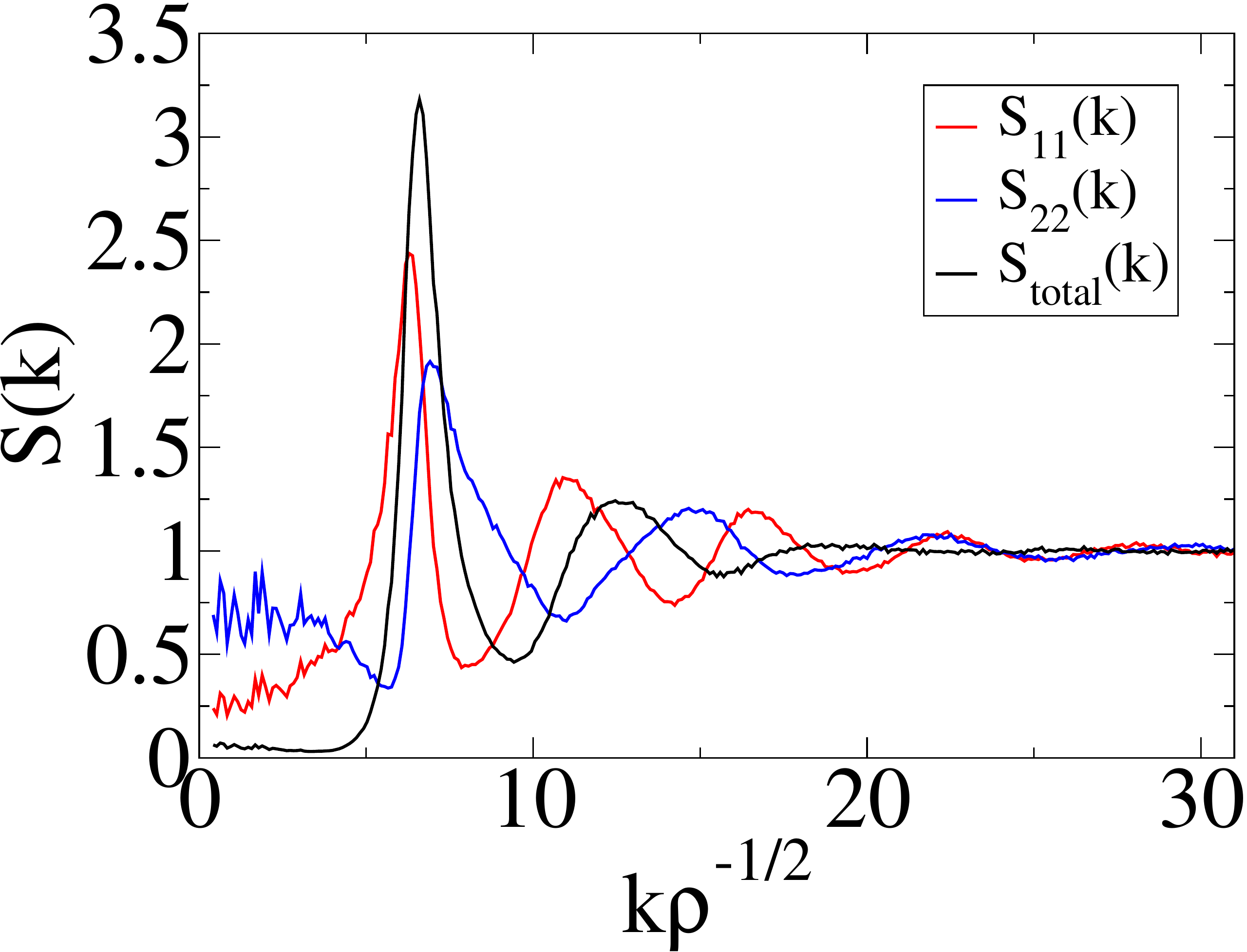}
}
\subfigure[\ $\Gamma_{22}=5$]{
\includegraphics[width=4cm, height=3cm]{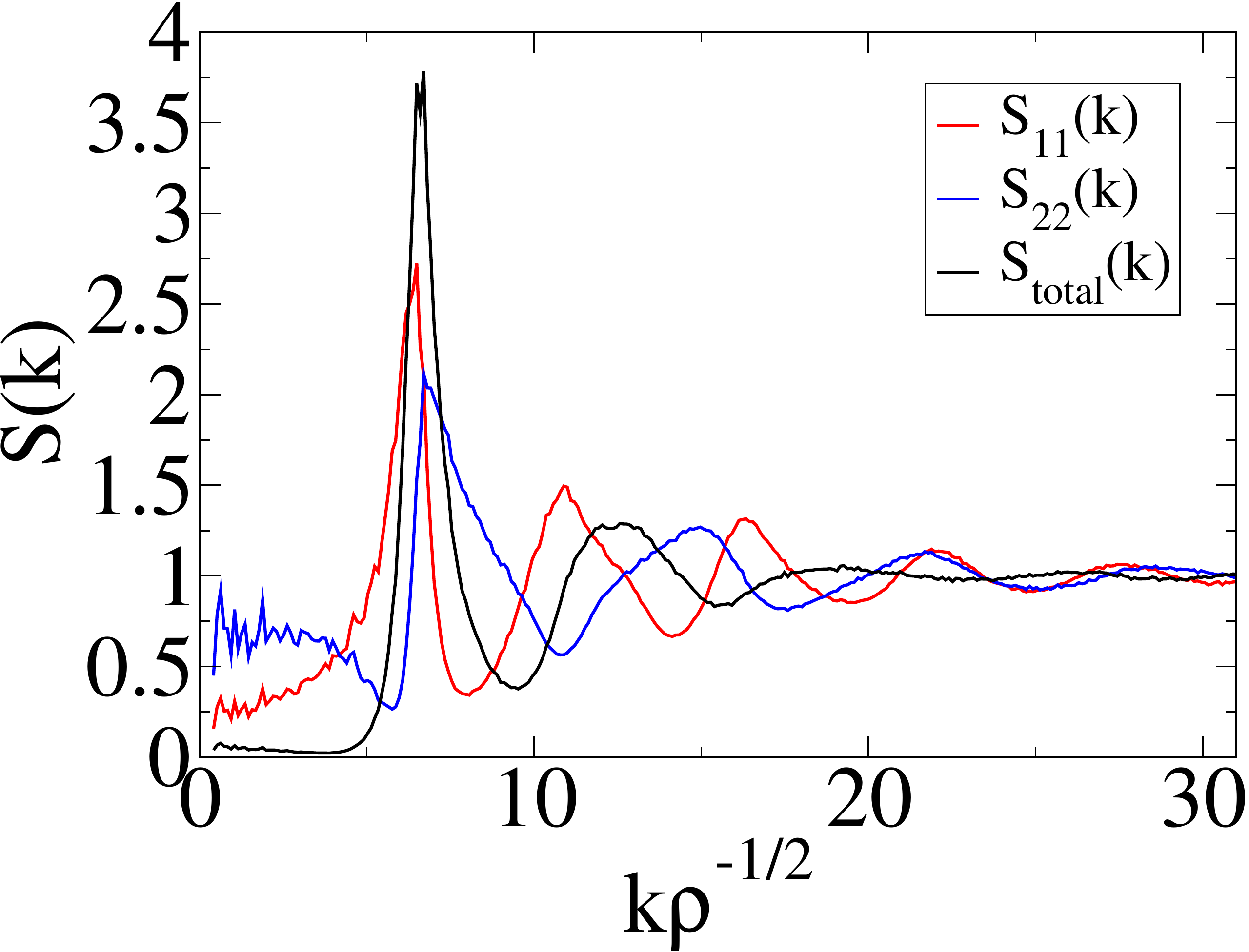}
}
\subfigure[\ $\Gamma_{22}=10$]{
\includegraphics[width=4cm, height=3cm]{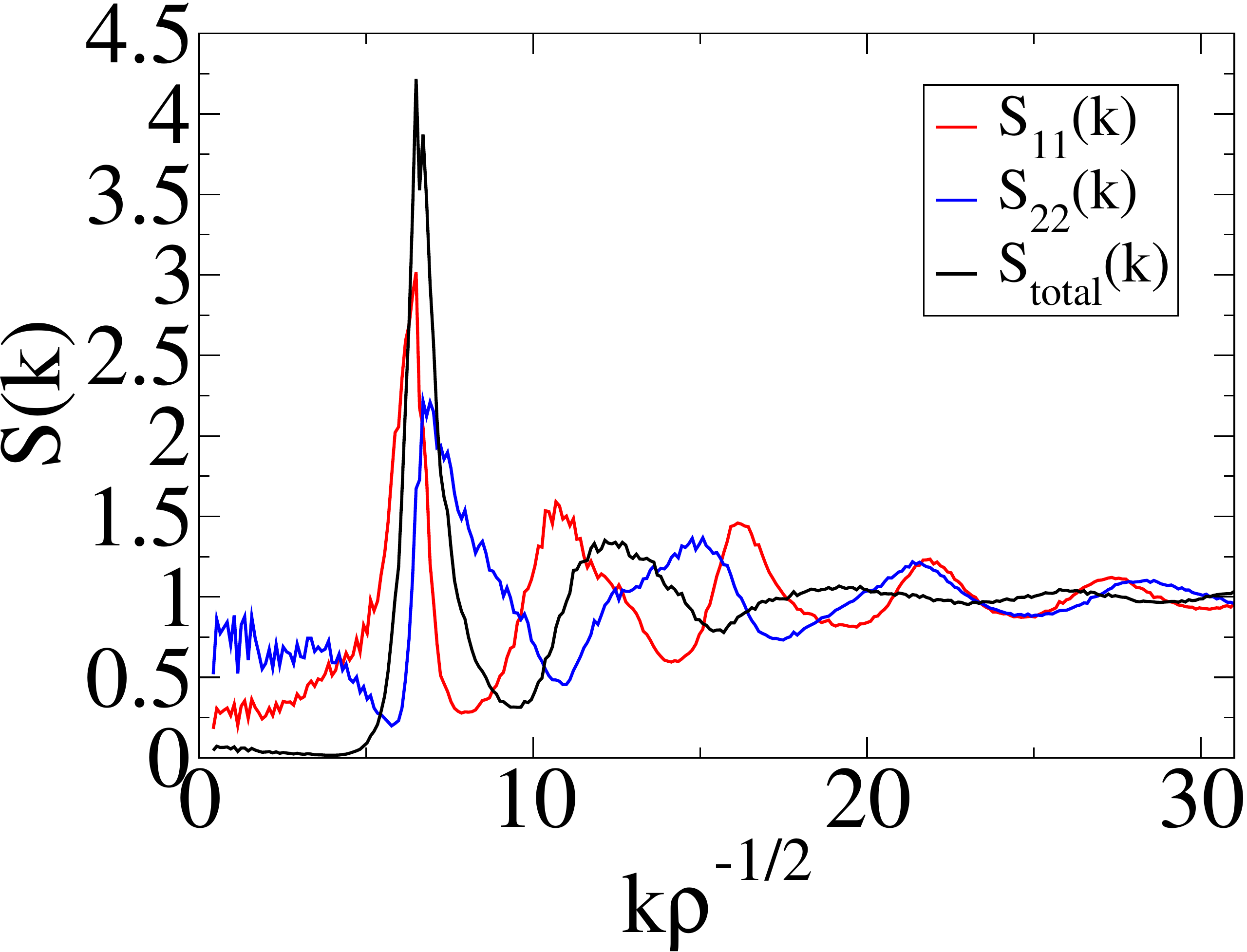}
}
\caption{Structure factors for large particles $S_{11}(k)$, small particles $S_{22}(k)$ and the entire system $S_{\text{total}}(k)$ for different values of the coupling strength $\Gamma_{22}$. Clearly, neither the point pattern associated with each component or the entire system is hyperuniform.}
\label{fig: bisk}
\end{figure}

However, we show that incompressibility in our equilibrium binary systems means the destructive interference scattering events between both species, rather than the vanishing of $S(k)$ itself as the case in monodisperse systems. To demonstrate this idea, we present simulation results of a binary system with the number density ratio $\rho_1/\rho_2=1$ and the susceptibility ratio $\chi_1/\chi_2=2$. Figure \ref{fig: bisk} depicts the computed structure factors for the small particles $S_{22}(k)$, large particles $S_{11}(k)$, and the entire system $S_{\text{\text{total}}}(k)$ for different coupling strengths. Here $S_{\text{\text{total}}}(k)$ can be written as $N^{-1}\left \langle |\tilde n_i(k) + \tilde n_j(k)|^2\right \rangle$.

\begin{figure}[H]
\centering
\subfigure[]{
\includegraphics[width=4.5cm, height=3.4cm]{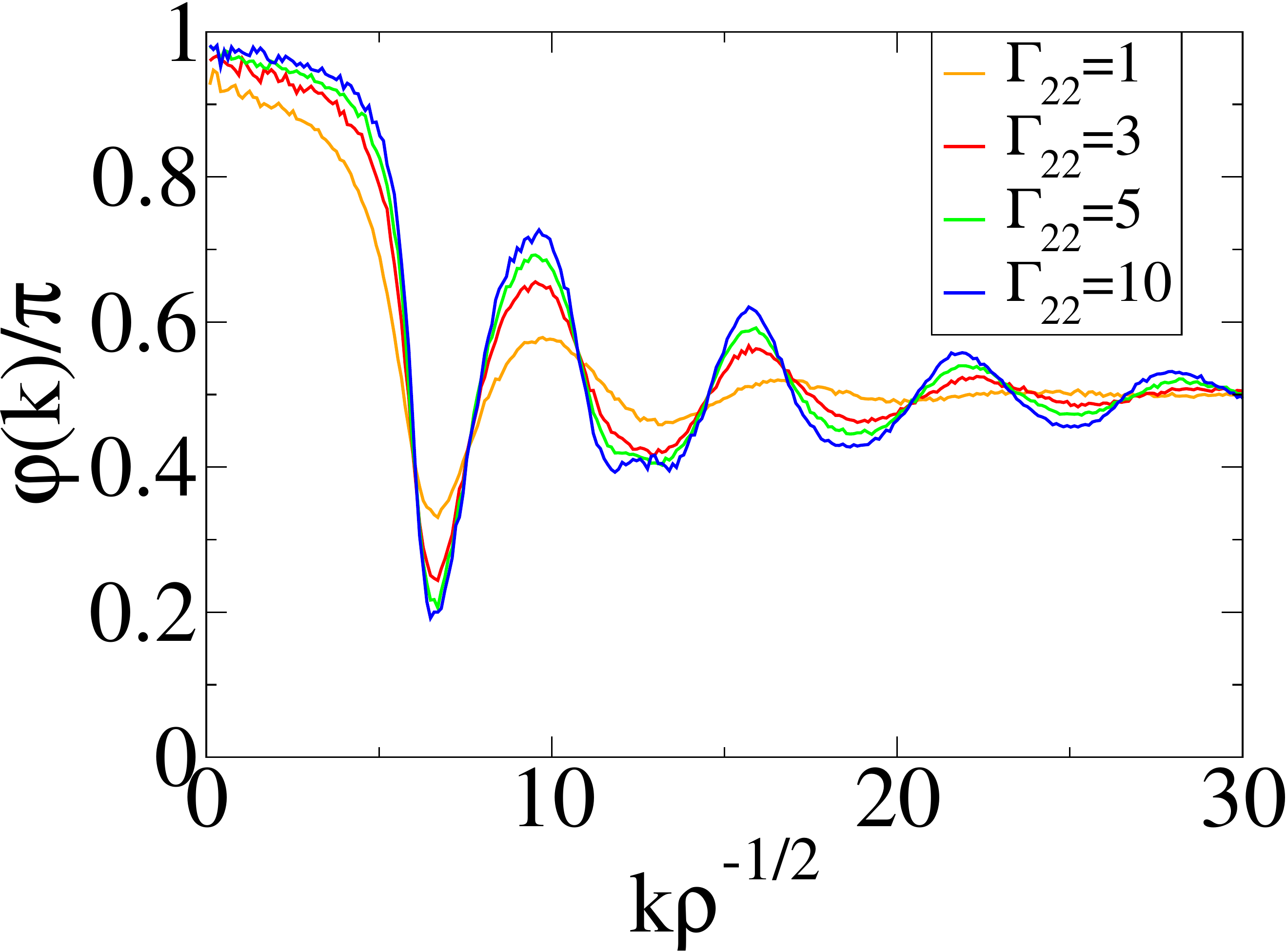}
}
\subfigure[]{
\includegraphics[width=3.6cm, height=3.15cm]{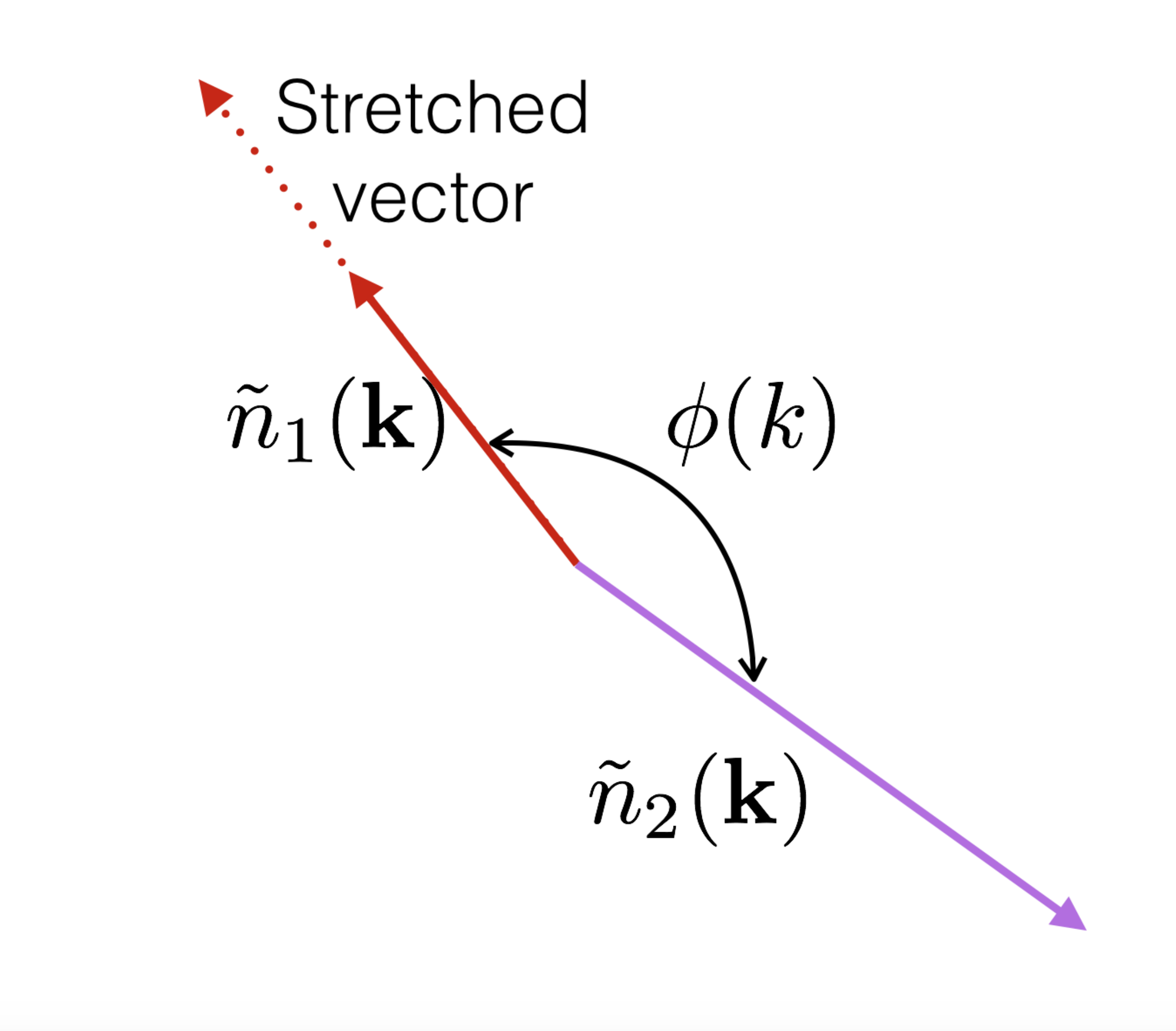}
}
\caption{
(a) The angle $\phi(k)$ between the complex collective
density variables for small particles $\tilde n_2(\mathbf k)$ and large particles $\tilde n_1(\mathbf k)$ as a function of wavenumber $k$ under different coupling strengths.
(b) A schematic plot of the complex collective density variables for small particles $\tilde n_2(\mathbf k)$ and large particles $\tilde n_1(\mathbf k)$ in the complex plane. Here, vector $\tilde n_1(\mathbf k)$ can be stretched out to cancel $\tilde n_2(\mathbf k)$ when they are added together.}
\label{fig: phase}
\end{figure}

\noindent Clearly, neither the point pattern associated with each component or the entire system of points is hyperuniform, which requires vanishing structure factors when $k$ goes to zero. Interestingly, we find that compared to the structure factors of each component, the small-$k$ values of the structure factor $S_{\text{total}}(k)$ are suppressed while the first peak is significantly enhanced. The $k$-value of the first
intersection of $S_{11}(k)$ and $S_{22}(k)$ corresponds to the minimum
of $S_{\text{total}}(k)$, which decreases as the coupling strength increases. These facts suggest that the interference due to
scattering from the small particles and large particles is constructive around the wavelength associated with the first
peak of the structure factor, while destructive in the small-$k$ region, which favors hyperuniformity. To demonstrate this observation
quantitatively, we compute the angle $\phi(\mathbf k)$ between $\tilde n_1(\mathbf k)$ and $\tilde n_2(\mathbf k)$, defined by
 \begin{equation} \label{def2}
 \phi(\mathbf k)=\arccos\frac{\operatorname{Re}(\tilde n_1(\mathbf k))\operatorname{Re}(\tilde n_2(\mathbf k))+\operatorname{Im}(\tilde n_1(\mathbf k))\operatorname{Im}(\tilde n_2(\mathbf k))}{|\tilde n_1(\mathbf k)||\tilde n_2(\mathbf k)|},
 \end{equation}
We plot the angular-averaged $\phi(k)$ as a function of wavenumber $k$ in Fig. \ref{fig: phase} (a). These results justify our previous arguments, i.e., in the small-$k$ region, the angle
$\phi(k)$ is very close to $\pi$, showing that the two complex
collective density variables align themselves in opposite directions and thus the
interference is destructive (see schematic shown in
Fig. \ref{fig: phase} (b)); while the first dip of $\phi(k)$ coincides
with the location of the first peak, showing that the interference is
constructive at the peak. When $k \rightarrow \infty$, the angle converges to
${\pi}/{2}$, which confirms the expectation that the correlation finally dies
out. Importantly, as the coupling strength $\Gamma_{22}$ increases, the
two complex collective density variables are more strongly aligned
with each other. We now show how this finding relates to Eq. (\ref{S_12}). Using the aforementioned complex collective density variables, it easily follows that $S_{12}(k)$ can be written as $\sqrt{S_{11}(k)S_{22}(k)}\cos{\phi(k)}$. Plugging into Eq. (\ref{S_12}), we have
\begin{equation} \label{compress_phi}
S_{11}(0^{+})S_{22}(0^{+})[1-\cos^2{\phi(0^{+})}]=0.
\end{equation}
The solution of Eq. (\ref{compress_phi}) means that $\phi(0^{+})=\pi$. Moreover, for dense liquids, $S_{11}(0^{+})$ and $S_{22}(0^{+})$ are insensitive to $\Gamma$ and thus we approximately have $(1+ \cos{\phi(0^{+})})\propto
T\kappa_T$. This relation explains the results in Fig. \ref{fig: phase} (b), i.e., the angle $\phi(0^{+})$ converges to $\pi$ as the coupling strength increases. 

However, although $\tilde n_1(0^{+})$ and $\tilde n_2(0^{+})$ almost align themselves
in opposite directions in the complex plane, due to the fact that their magnitudes are
different, their sum as well as the resulting \text{total} structure factor $S_{\text{total}}(0^{+})$ can
not vanish. This suggests that if we can elongate the vectors $\tilde n_1(0^{+})$
and $\tilde n_2(0^{+})$ such that their magnitudes match each other,
their sum would be very close to zero. Importantly, this idea enables us to design
an optimized hyperuniform two-phase system based on decorating the point pattern. 

For a two-phase medium, hyperuniformity is defined in terms
of the spectral density $\tilde\chi_{_V}(\mathbf k)$, as noted earlier. In general, the spectral density can be written as $|\tilde J(k)|^2/V$, where $\tilde J(k)$ is the Fourier transform of $J(\mathbf x)=I(\mathbf x)-\left\langle I(\mathbf x)\right\rangle$ and $I(\mathbf x)$ is the indicator function for the particle phase \cite{PhysRevE.94.022122}. In the small-$k$ region, we have the following approximate expression for our binary system  
\begin{equation} \label{Ik}
\tilde J(k)=\pi R_1^2\tilde n_1(k)+\pi R_2^2\tilde n_2(k).
\end{equation}   
To make $|\tilde J(0^{+})|$ as close to zero as possible, we immediately come to the relation that 
$
R_1^2|\tilde n_1(0^{+})|= R_2^2|\tilde n_2(0^{+})|,
$ 
which leads to the optimal particle size ratio 
\begin{equation} \label{ratio}
\frac{R_1}{R_2}=\left(\frac{\rho_2S_{22}(0^{+})}{\rho_1S_{11}(0^{+})}\right)^{\frac{1}{4}}.
\end{equation} 
Generally, the susceptibility ratio $\chi_1/\chi_2$ is only dependent on the doping level, thus can be 
independent of the physical size ratio $R_1/R_2$. This provides great
flexibility to tune the system to the optimal hyperuniform state. Observe that as the coupling strength increases, the values
of the structure factors $S_{11}(0^{+})$ and $S_{22}(0^{+})$
approximately remains the same. This insensitivity shows that the optimal particle size ratio is
basically determined by the susceptibility ratio $\chi_1/\chi_2$ and
the number density ratio $\rho_1/\rho_2$. This is particularly
important from an experimental point of view because it means that
the optimal colloid composition can be prescribed and one only needs to tune the magnetic field to the desired level.  

\begin{figure}[]
\centering
\subfigure[\ $\Gamma_{22}=1$]{
\includegraphics[width=4cm, height=2.6cm]{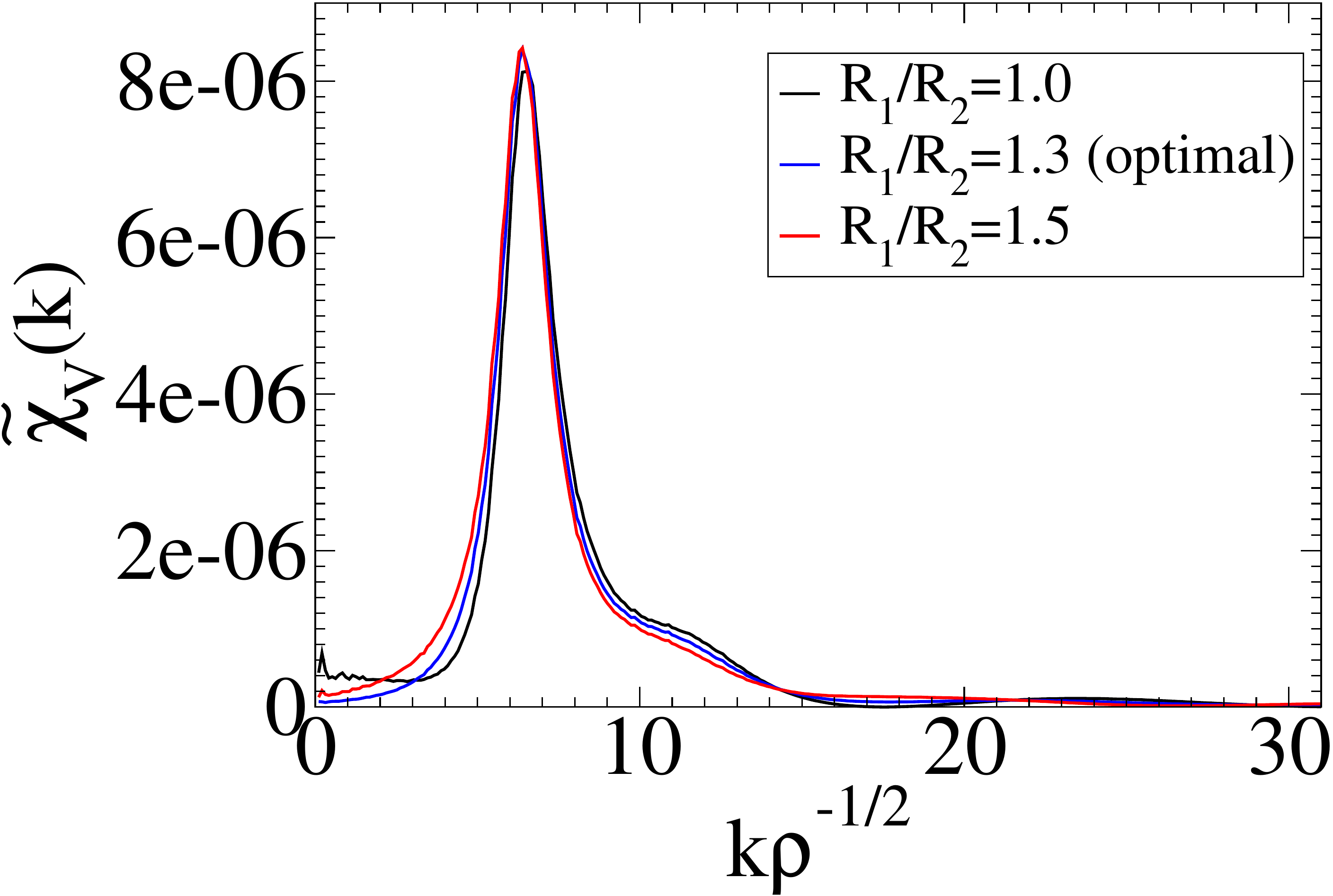}
}
\subfigure[\ $\Gamma_{22}=3$]{
\includegraphics[width=4cm, height=2.6cm]{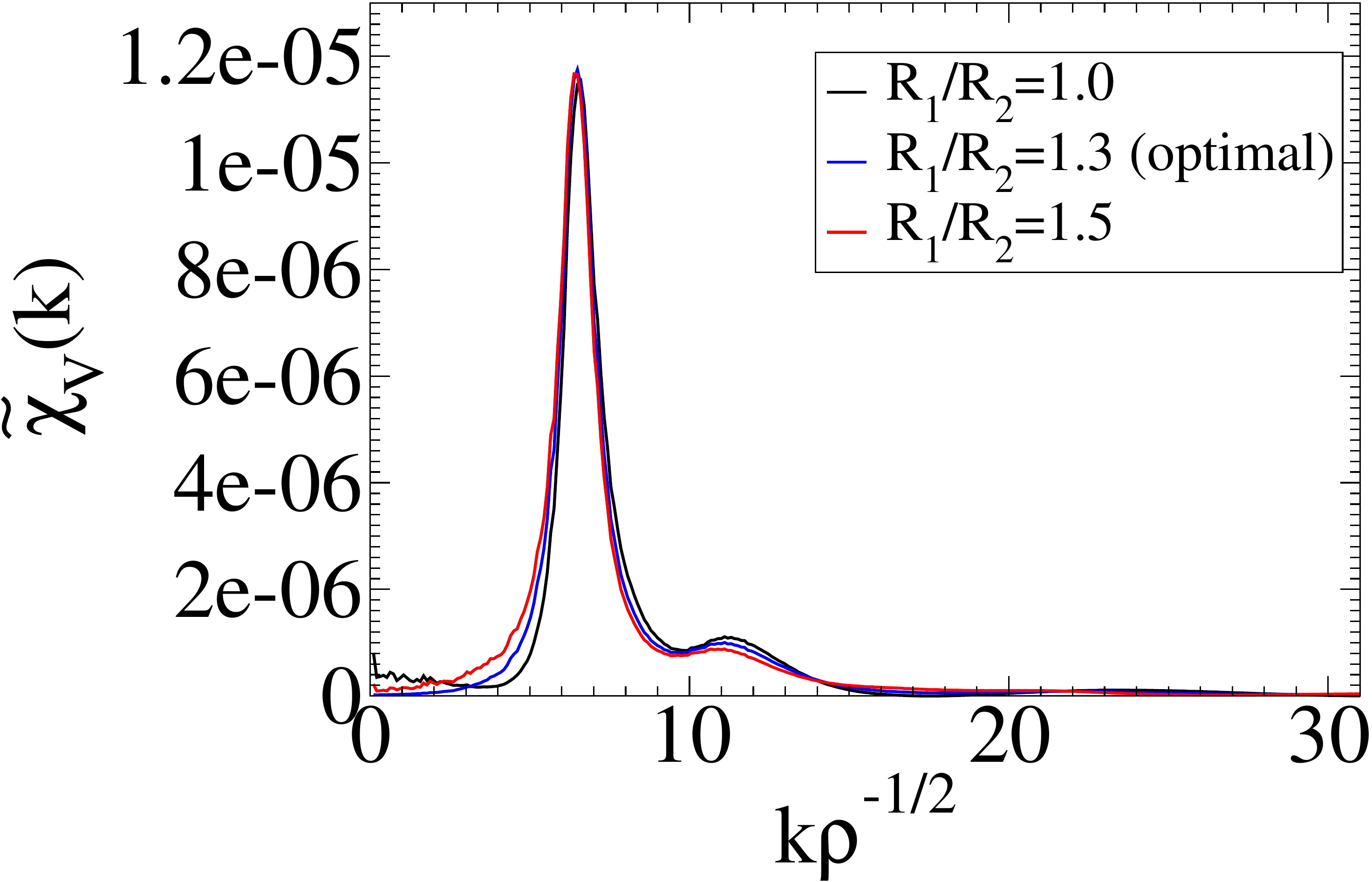}
}
\subfigure[\ $\Gamma_{22}=5$]{
\includegraphics[width=4cm, height=2.6cm]{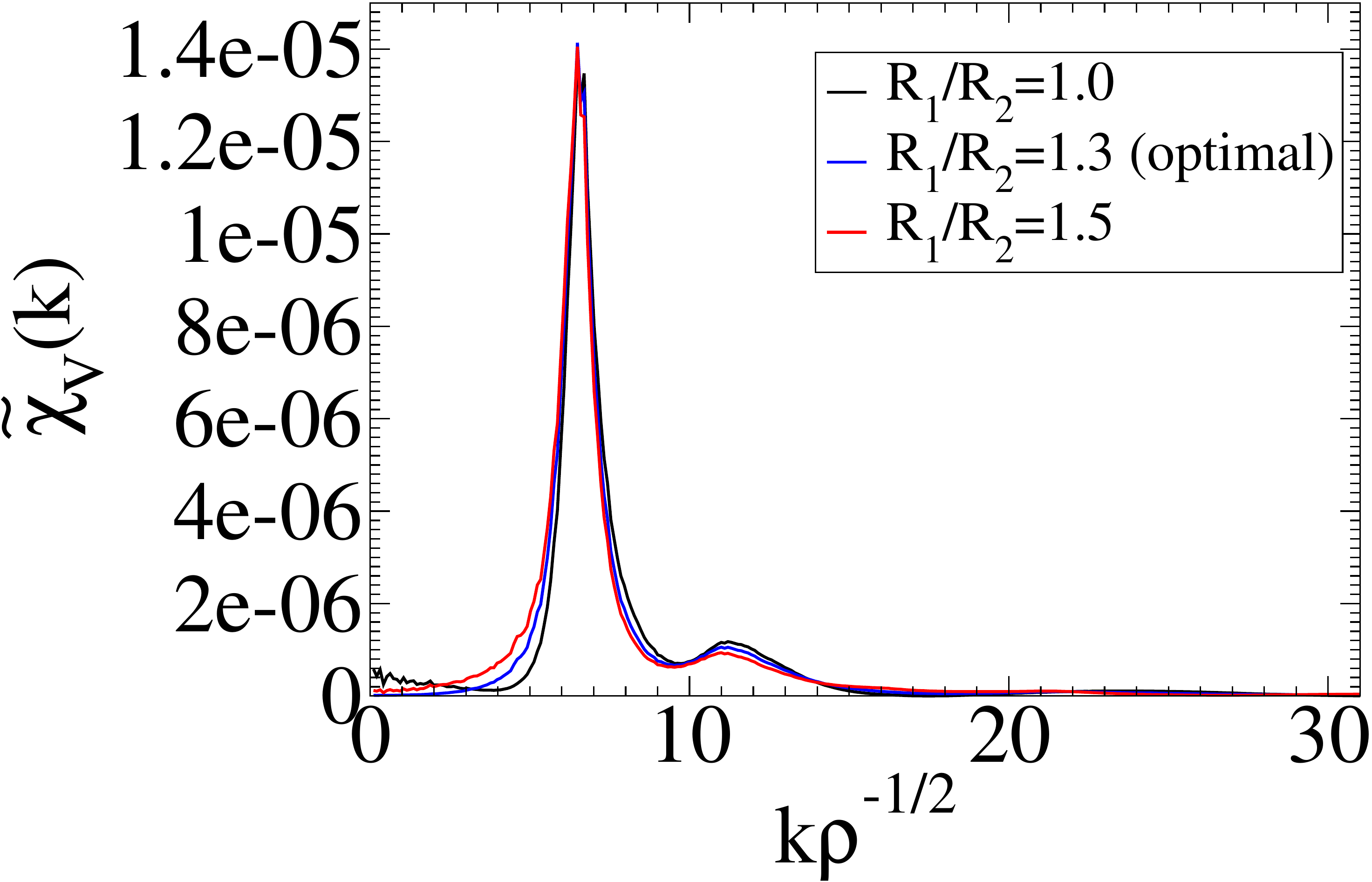}
}
\subfigure[\ $\Gamma_{22}=10$]{
\includegraphics[width=4cm, height=2.6cm]{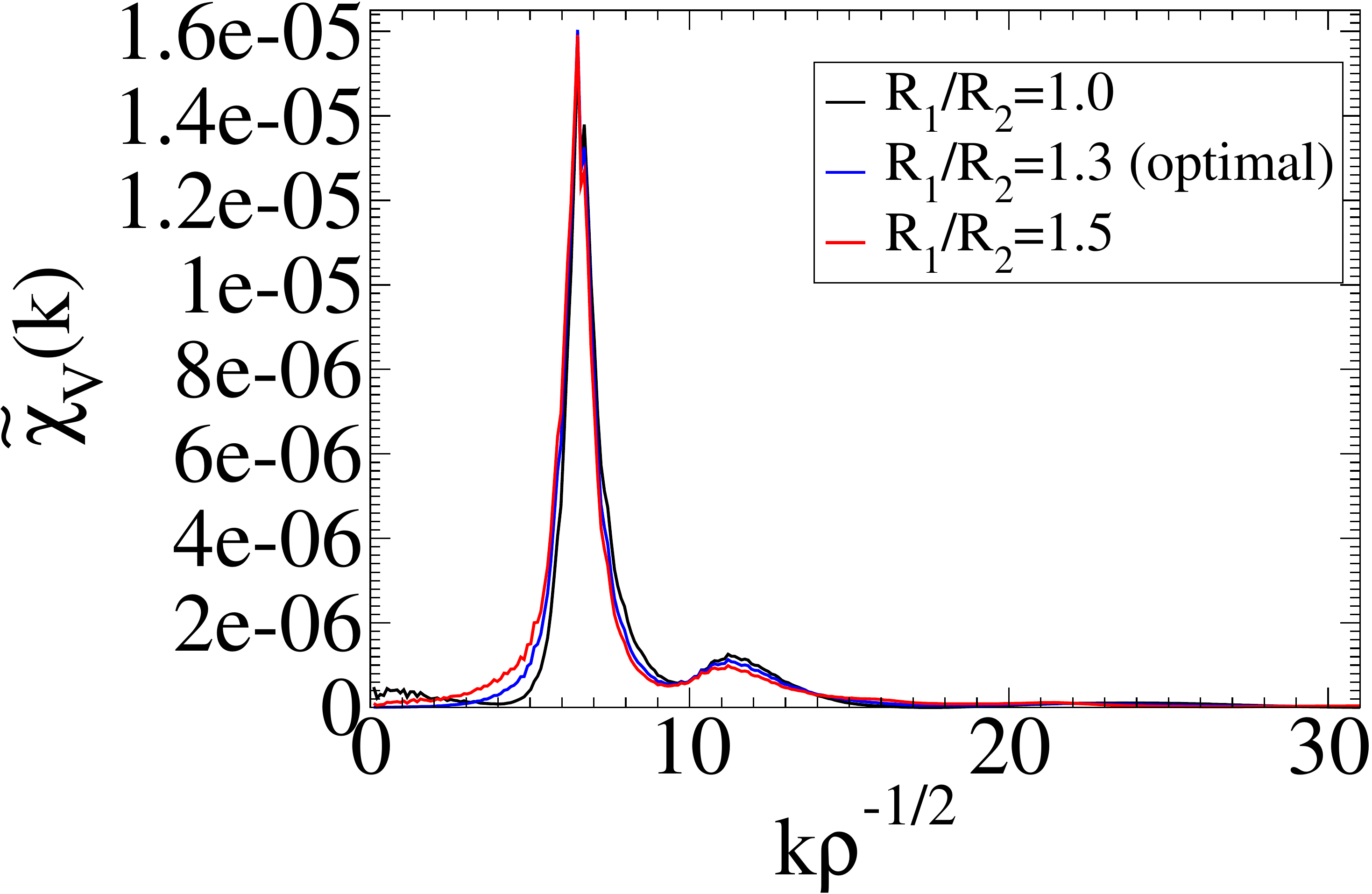}
}
\caption{Spectral densities $\tilde \chi_{_V}(k)$ for binary systems with different size ratios $R_1/R_2$. The volume fraction of the particle phase is fixed at 0.15.} 
\label{fig: chik}
\end{figure}

\begin{figure*}[]
\centering
\subfigure[]{
\includegraphics[width=4.5cm, height=4.5cm]{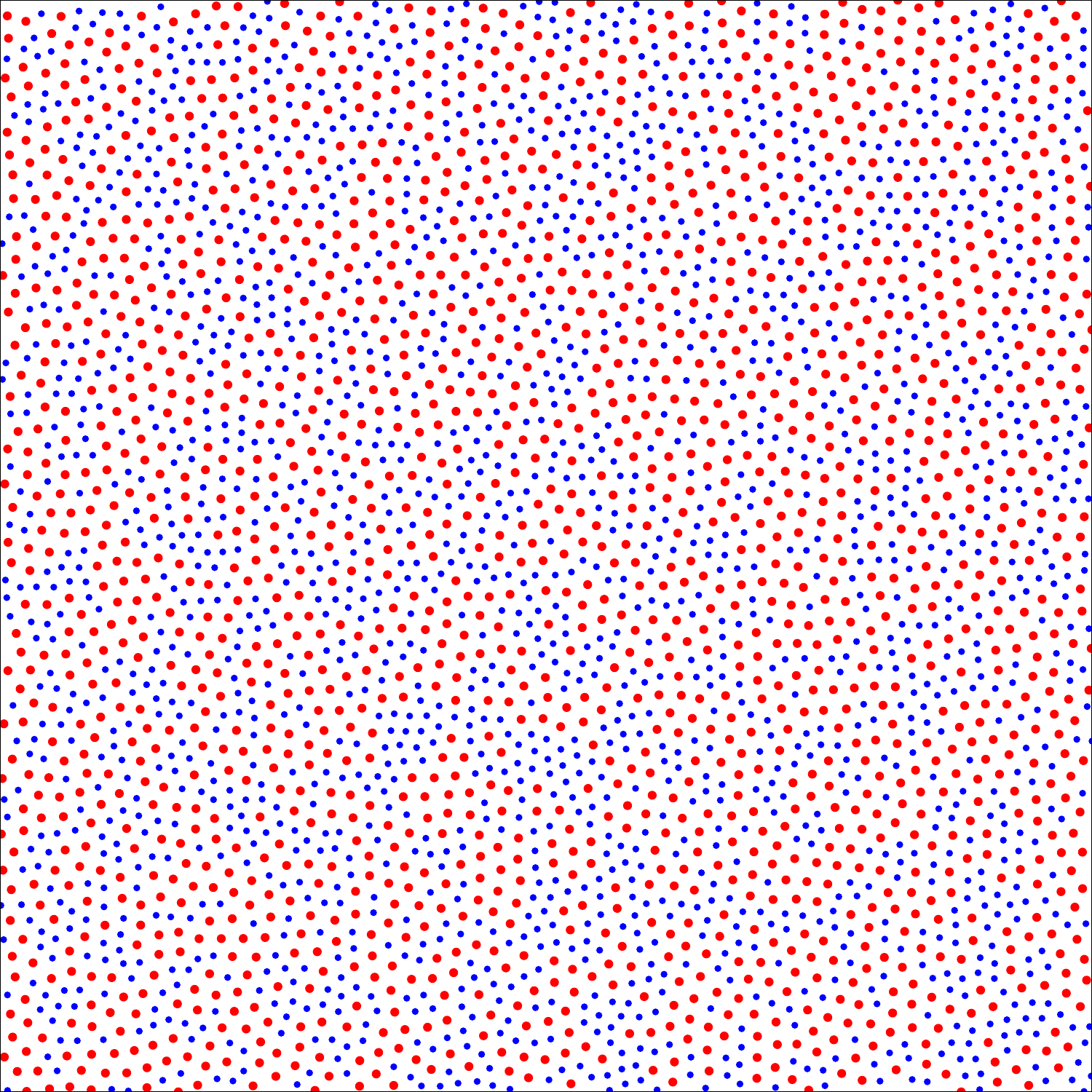}
}
\subfigure[]{
\includegraphics[width=6cm, height=4cm]{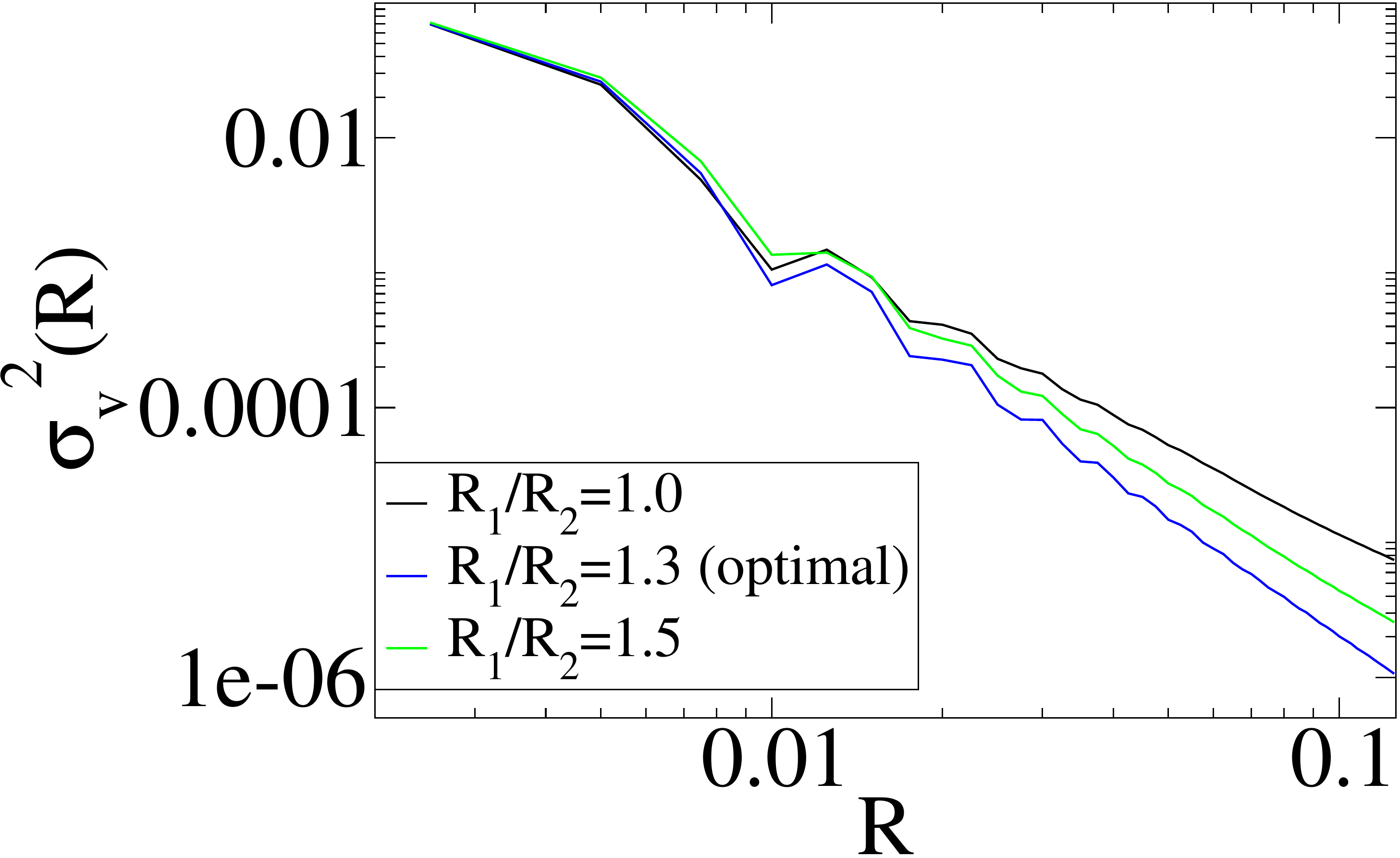}
}
\subfigure[]{
\includegraphics[width=6cm, height=4.5cm]{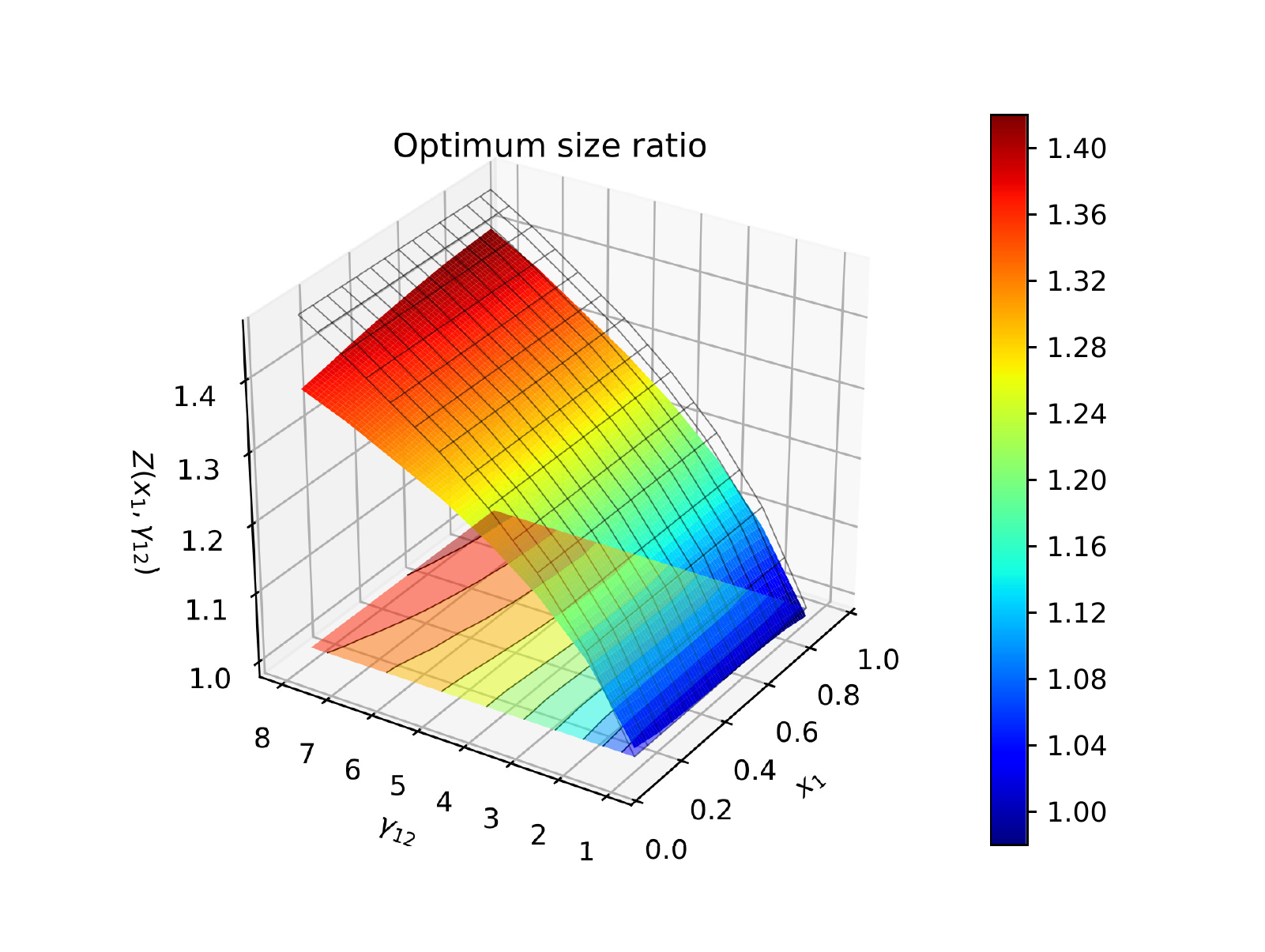}
}
\caption{(a) A realization of the optimal two-phase system at $\Gamma_{22}=5$ with volume fraction 0.15. (b) The local volume fraction variance $\sigma_V^2(R)$ as a function of the spherical observation window of radius $R$ for different size ratios (fixed volume fraction). (c) Optimal size ratio $Z(x_1,\gamma_{12})$ in terms of composition and coupling strength ratio, $\gamma_{12}$, as defined by Eq. (\ref{ratio}).}
\label{fig: var}
\end{figure*}

To demonstrate the effectiveness of our protocol, we consider decorating the simulated point pattern with particles with different size ratios while keep the volume fraction of the particle phase fixed (here we use 0.15). Figure \ref{fig: chik} shows corresponding spectral densities $\tilde \chi_{_V}(k)$ for these systems. Indeed, we find that the optimal size
ratio ($\approx 1.3$) gives the smallest $\tilde \chi_{_V}(0^{+})$. To quantify how
close the system is to perfect hyperuniformity, we employ the
``hyperuniformity index", $H$, defined as 
$
H={\tilde \chi_{_V}(0^{+})}/{\tilde \chi_{_V}(k_{max})},
$
where $\tilde \chi_{_V}(k_{max})$ is the value of the largest peak of
the spectral density \cite{atkinson2016critical}. We find for $\Gamma_{22}=10$, the
``hyperuniformity index" $H$ is as small as 0.0004, which is two orders of magnitude smaller than the value obtained from the system with identical particle sizes. Moreover, we find
empirically that for the optimal structure, the relation $\tilde
\chi_{_V}(0^{+}) \propto 1/\Gamma_{22}$ holds, which is very similar
to the behavior of the monodisperse structure factors. This interesting similarity between the spectral density $\tilde \chi_{_V}(k)$ of the binary system under the optimal size ratio and the structure factor in monodisperse systems is straightforward to demonstrate. Specifically, plugging in the optimal size ratio into Eq. (\ref{Ik}) we find $\tilde \chi_{_V}(0^{+}) \propto 2\pi^2 R_1^4|\tilde n_1(0^{+})|^2[1+ \cos{\phi(0^{+})}]\propto T\kappa_T$. In the case of dense dipolar systems, this implies that we approximately have $\tilde \chi_{_V}(0^{+}) \propto 1/\Gamma$ \cite{khrapak2018thermodynamics}, which indeed is what we find in simulations. This finding suggests that we can control the volume fraction fluctuations at long wavelengths by tuning the magnetic
field, while retaining the isotropy of a disordered system. We depict a realization of the optimal system at
$\Gamma_{22}=5$ in Fig. \ref{fig: var} (a). In Fig. \ref{fig: var} (b), we directly compute the local volume fraction variances $\sigma_V^2(R)$ associated with windows of radii $R$ \cite{PhysRevE.94.022122} for different size ratios. Note that at the
optimal size ratio ($\approx 1.3$), the volume fraction variance indeed decreases fastest with the increase of $R$. Moreover, a scaling of $R^{-2.9}$ is found for the optimal configuration. These results further confirm that our protocol gives the most hyperuniform configuration among all possible size ratios.

Similar results can be found for other set of parameters. To provide a useful recipe, we use the RY approximation to systematically study the manifold of the optimal size ratio as a function of the coupling strength ratio,
$\gamma_{12}=\Gamma_{11}/\Gamma_{22}$ and composition. The function
$Z(x_1,\gamma_{12})=R_1/R_2$ is computed from
Eq. (\ref{ratio}) where the partial structure factors are determined
using the RY approximation. This quantity is plotted in Fig. \ref{fig: var} (c). Interestingly, the optimal size ratio is almost independent of the composition as long as $\gamma_{12}$ is not too large. Using the effective hard disk diameter  
$\int_{0}^{\infty}(1-\exp(-{u(r)}/{k_BT}))dr \approx {1.354\Gamma^{\frac{1}{3}}}/{\sqrt{\rho_1}}$, we find a good empirical expression for the optimal ratio in this region:
$Z(x_1,\gamma_{12}) \sim ({\gamma_{12}}^{\frac{1}{3}}+{\gamma_{12}}^{\frac{1}{6}}+1)/3$, which is represented by the wireframe in Fig. \ref{fig: var} (c). Using this figure, one may determine the experimental characteristics for our colloidal mixture
to become effectively hyperuniform.

{\it Generalizations}. 
Our protocol can be applied to other long-ranged interactions, dimensionalities and/or polydispersity.  
Recall that to reach the key relation $\phi(k) \approx \pi$, we have not assumed anything about the exact form of the interaction, nor dimensionality, as long as the system is hard to compress. For polydisperse systems, incompressibility means that there exists an eigenvector of the structure factor matrix
$S_{ij}$ which leads to a vanishing eigenvalue at infinite
wavelength \cite{berthier2011suppressed}. Thus there exists a vector $\mathbf a$ such that  
\begin{equation} \label{eigen}
\sum_{i,j} a_iS_{ij}(0^{+})a_j=0. 
\end{equation}
To make the polydisperse system a hyperuniform two-phase medium by decoration, note
that the spectral density $\tilde \chi_{_V}(0^{+})$ has the following form 
\begin{equation} \label{chiv_poly}
\tilde \chi_{_V}(0^{+}) \propto \sum_{i,j} \sqrt{x_i}V(R_i)S_{ij}(0^{+})\sqrt{x_j}V(R_j),  
\end{equation}
where $V(R_i)$ is the volume of a particle with radius $R_i$. Comparing Eq. (\ref{eigen}) and Eq. (\ref{chiv_poly}) it immediately follows that the optimal composition has the property that $V(R_i) \propto a_i/\sqrt{x_i}$, i.e., $R_i \propto a_i^{1/d}/{x_i}^{1/2d}$. For the dipolar system we consider, it is easy to verify that $a=(1/\sqrt{S_{11}(0^{+})}, 1/\sqrt{S_{22}(0^{+})})$ is the eigenvector that we are looking for. Plug in $a$ and $d=2$, we indeed recover the result in Eq. (\ref{ratio}). Furthermore, we note that using the same protocol we can make other spatial variables hyperuniform, e.g., by choosing ${R_1}/{R_2}=({\rho_2S_{22}(0^{+})}/{\rho_1S_{11}(0^{+})})^{\frac{1}{2}}$ the binary system is hyperuniform with respect to surface area fluctuations; see Supplemental Material \cite{supplement}.  

In conclusion, we propose a highly feasible and robust equilibrium protocol that can be employed in the laboratory to fabricate large disordered hyperuniform materials, using binary superparamagnetic colloidal particles confined in a 2D plane. The destructive interference of scattering events between the two species of the binary system at infinite wavelength enables us to design the optimal size ratio. Although the present work stresses 2D dipolar binary systems, our protocol turns out to be general and suitable to
systems with other sufficiently long-ranged soft interactions, dimensionalities and/or polydispersity. Our methodology opens up avenues to synthesize large, tunable photonic hyperuniform materials that function in the visible to infrared range and thus may accelerate the discovery of novel photonic materials.

\begin{acknowledgements}
\indent The authors are grateful to C. Maher, T. Middlemas and C. Likos for fruitful discussions. Z.M. and S.T. acknowledge the support of the National Science Foundation under Grant No. DMR-1714722. EL acknowledges the support from the Agencia Estatal de Investigación and Fondo Europeo de Desarrollo Regional (FEDER) under grant No. FIS2017-89361-C3-2-P and from European Union’s Horizon 2020 Research and Innovation Staff Exchange programme under the Marie  Skłodowska-Curie grant agreement No 734276.
\end{acknowledgements}


\begin{thebibliography}{52}%
\makeatletter
\providecommand \@ifxundefined [1]{%
 \@ifx{#1\undefined}
}%
\providecommand \@ifnum [1]{%
 \ifnum #1\expandafter \@firstoftwo
 \else \expandafter \@secondoftwo
 \fi
}%
\providecommand \@ifx [1]{%
 \ifx #1\expandafter \@firstoftwo
 \else \expandafter \@secondoftwo
 \fi
}%
\providecommand \natexlab [1]{#1}%
\providecommand \enquote  [1]{``#1''}%
\providecommand \bibnamefont  [1]{#1}%
\providecommand \bibfnamefont [1]{#1}%
\providecommand \citenamefont [1]{#1}%
\providecommand \href@noop [0]{\@secondoftwo}%
\providecommand \href [0]{\begingroup \@sanitize@url \@href}%
\providecommand \@href[1]{\@@startlink{#1}\@@href}%
\providecommand \@@href[1]{\endgroup#1\@@endlink}%
\providecommand \@sanitize@url [0]{\catcode `\\12\catcode `\$12\catcode
  `\&12\catcode `\#12\catcode `\^12\catcode `\_12\catcode `\%12\relax}%
\providecommand \@@startlink[1]{}%
\providecommand \@@endlink[0]{}%
\providecommand \url  [0]{\begingroup\@sanitize@url \@url }%
\providecommand \@url [1]{\endgroup\@href {#1}{\urlprefix }}%
\providecommand \urlprefix  [0]{URL }%
\providecommand \Eprint [0]{\href }%
\providecommand \doibase [0]{http://dx.doi.org/}%
\providecommand \selectlanguage [0]{\@gobble}%
\providecommand \bibinfo  [0]{\@secondoftwo}%
\providecommand \bibfield  [0]{\@secondoftwo}%
\providecommand \translation [1]{[#1]}%
\providecommand \BibitemOpen [0]{}%
\providecommand \bibitemStop [0]{}%
\providecommand \bibitemNoStop [0]{.\EOS\space}%
\providecommand \EOS [0]{\spacefactor3000\relax}%
\providecommand \BibitemShut  [1]{\csname bibitem#1\endcsname}%
\let\auto@bib@innerbib\@empty
\bibitem [{\citenamefont {Torquato}\ and\ \citenamefont
  {Stillinger}(2003)}]{torquato2003local}%
  \BibitemOpen
  \bibfield  {author} {\bibinfo {author} {\bibfnamefont {S.}~\bibnamefont
  {Torquato}}\ and\ \bibinfo {author} {\bibfnamefont {F.~H.}\ \bibnamefont
  {Stillinger}},\ }\href@noop {} {\bibfield  {journal} {\bibinfo  {journal}
  {Phys. Rev. E}\ }\textbf {\bibinfo {volume} {68}},\ \bibinfo {pages} {041113}
  (\bibinfo {year} {2003})}\BibitemShut {NoStop}%
\bibitem [{\citenamefont {Dyson}(1962)}]{dyson1962statistical}%
  \BibitemOpen
  \bibfield  {author} {\bibinfo {author} {\bibfnamefont {F.~J.}\ \bibnamefont
  {Dyson}},\ }\href@noop {} {\bibfield  {journal} {\bibinfo  {journal} {J.
  Math. Phys}\ }\textbf {\bibinfo {volume} {3}},\ \bibinfo {pages} {140}
  (\bibinfo {year} {1962})}\BibitemShut {NoStop}%
\bibitem [{\citenamefont {Jancovici}(1981)}]{jancovici1981exact}%
  \BibitemOpen
  \bibfield  {author} {\bibinfo {author} {\bibfnamefont {B.}~\bibnamefont
  {Jancovici}},\ }\href@noop {} {\bibfield  {journal} {\bibinfo  {journal}
  {Phys. Rev. Lett.}\ }\textbf {\bibinfo {volume} {46}},\ \bibinfo {pages}
  {386} (\bibinfo {year} {1981})}\BibitemShut {NoStop}%
\bibitem [{\citenamefont {Torquato}\ \emph {et~al.}(2015)\citenamefont
  {Torquato}, \citenamefont {Zhang},\ and\ \citenamefont
  {Stillinger}}]{torquato2015ensemble}%
  \BibitemOpen
  \bibfield  {author} {\bibinfo {author} {\bibfnamefont {S.}~\bibnamefont
  {Torquato}}, \bibinfo {author} {\bibfnamefont {G.}~\bibnamefont {Zhang}}, \
  and\ \bibinfo {author} {\bibfnamefont {F.~H.}\ \bibnamefont {Stillinger}},\
  }\href@noop {} {\bibfield  {journal} {\bibinfo  {journal} {Phys. Rev. X}\
  }\textbf {\bibinfo {volume} {5}},\ \bibinfo {pages} {021020} (\bibinfo {year}
  {2015})}\BibitemShut {NoStop}%
\bibitem [{\citenamefont {Donev}\ \emph {et~al.}(2005)\citenamefont {Donev},
  \citenamefont {Stillinger},\ and\ \citenamefont
  {Torquato}}]{PhysRevLett.95.090604}%
  \BibitemOpen
  \bibfield  {author} {\bibinfo {author} {\bibfnamefont {A.}~\bibnamefont
  {Donev}}, \bibinfo {author} {\bibfnamefont {F.~H.}\ \bibnamefont
  {Stillinger}}, \ and\ \bibinfo {author} {\bibfnamefont {S.}~\bibnamefont
  {Torquato}},\ }\href {\doibase 10.1103/PhysRevLett.95.090604} {\bibfield
  {journal} {\bibinfo  {journal} {Phys. Rev. Lett.}\ }\textbf {\bibinfo
  {volume} {95}},\ \bibinfo {pages} {090604} (\bibinfo {year}
  {2005})}\BibitemShut {NoStop}%
\bibitem [{\citenamefont {Kurita}\ and\ \citenamefont
  {Weeks}(2011)}]{kurita2011incompressibility}%
  \BibitemOpen
  \bibfield  {author} {\bibinfo {author} {\bibfnamefont {R.}~\bibnamefont
  {Kurita}}\ and\ \bibinfo {author} {\bibfnamefont {E.~R.}\ \bibnamefont
  {Weeks}},\ }\href@noop {} {\bibfield  {journal} {\bibinfo  {journal} {Phys.
  Rev. E}\ }\textbf {\bibinfo {volume} {84}},\ \bibinfo {pages} {030401}
  (\bibinfo {year} {2011})}\BibitemShut {NoStop}%
\bibitem [{\citenamefont {Zachary}\ \emph {et~al.}(2011)\citenamefont
  {Zachary}, \citenamefont {Jiao},\ and\ \citenamefont
  {Torquato}}]{zachary2011hyperuniformity}%
  \BibitemOpen
  \bibfield  {author} {\bibinfo {author} {\bibfnamefont {C.~E.}\ \bibnamefont
  {Zachary}}, \bibinfo {author} {\bibfnamefont {Y.}~\bibnamefont {Jiao}}, \
  and\ \bibinfo {author} {\bibfnamefont {S.}~\bibnamefont {Torquato}},\
  }\href@noop {} {\bibfield  {journal} {\bibinfo  {journal} {Phys. Rev. E}\
  }\textbf {\bibinfo {volume} {83}},\ \bibinfo {pages} {051308} (\bibinfo
  {year} {2011})}\BibitemShut {NoStop}%
\bibitem [{\citenamefont {Gerasimenko}\ \emph {et~al.}(2019)\citenamefont
  {Gerasimenko}, \citenamefont {Vaskivskyi}, \citenamefont {Litskevich},
  \citenamefont {Ravnik}, \citenamefont {Vodeb}, \citenamefont {Diego},
  \citenamefont {Kabanov},\ and\ \citenamefont
  {Mihailovic}}]{gerasimenko2019quantum}%
  \BibitemOpen
  \bibfield  {author} {\bibinfo {author} {\bibfnamefont {Y.~A.}\ \bibnamefont
  {Gerasimenko}}, \bibinfo {author} {\bibfnamefont {I.}~\bibnamefont
  {Vaskivskyi}}, \bibinfo {author} {\bibfnamefont {M.}~\bibnamefont
  {Litskevich}}, \bibinfo {author} {\bibfnamefont {J.}~\bibnamefont {Ravnik}},
  \bibinfo {author} {\bibfnamefont {J.}~\bibnamefont {Vodeb}}, \bibinfo
  {author} {\bibfnamefont {M.}~\bibnamefont {Diego}}, \bibinfo {author}
  {\bibfnamefont {V.}~\bibnamefont {Kabanov}}, \ and\ \bibinfo {author}
  {\bibfnamefont {D.}~\bibnamefont {Mihailovic}},\ }\href@noop {} {\bibfield
  {journal} {\bibinfo  {journal} {Nat. Mater}\ }\textbf {\bibinfo {volume}
  {18}},\ \bibinfo {pages} {1078} (\bibinfo {year} {2019})}\BibitemShut
  {NoStop}%
\bibitem [{\citenamefont {Hexner}\ and\ \citenamefont
  {Levine}(2015)}]{PhysRevLett.114.110602}%
  \BibitemOpen
  \bibfield  {author} {\bibinfo {author} {\bibfnamefont {D.}~\bibnamefont
  {Hexner}}\ and\ \bibinfo {author} {\bibfnamefont {D.}~\bibnamefont
  {Levine}},\ }\href {\doibase 10.1103/PhysRevLett.114.110602} {\bibfield
  {journal} {\bibinfo  {journal} {Phys. Rev. Lett.}\ }\textbf {\bibinfo
  {volume} {114}},\ \bibinfo {pages} {110602} (\bibinfo {year}
  {2015})}\BibitemShut {NoStop}%
\bibitem [{\citenamefont {Weijs}\ \emph {et~al.}(2015)\citenamefont {Weijs},
  \citenamefont {Jeanneret}, \citenamefont {Dreyfus},\ and\ \citenamefont
  {Bartolo}}]{weijs2015emergent}%
  \BibitemOpen
  \bibfield  {author} {\bibinfo {author} {\bibfnamefont {J.~H.}\ \bibnamefont
  {Weijs}}, \bibinfo {author} {\bibfnamefont {R.}~\bibnamefont {Jeanneret}},
  \bibinfo {author} {\bibfnamefont {R.}~\bibnamefont {Dreyfus}}, \ and\
  \bibinfo {author} {\bibfnamefont {D.}~\bibnamefont {Bartolo}},\ }\href@noop
  {} {\bibfield  {journal} {\bibinfo  {journal} {Phys. Rev. Lett.}\ }\textbf
  {\bibinfo {volume} {115}},\ \bibinfo {pages} {108301} (\bibinfo {year}
  {2015})}\BibitemShut {NoStop}%
\bibitem [{\citenamefont {Bertrand}\ \emph {et~al.}(2019)\citenamefont
  {Bertrand}, \citenamefont {Chatenay},\ and\ \citenamefont
  {Voituriez}}]{bertrand2019nonlinear}%
  \BibitemOpen
  \bibfield  {author} {\bibinfo {author} {\bibfnamefont {T.}~\bibnamefont
  {Bertrand}}, \bibinfo {author} {\bibfnamefont {D.}~\bibnamefont {Chatenay}},
  \ and\ \bibinfo {author} {\bibfnamefont {R.}~\bibnamefont {Voituriez}},\
  }\href@noop {} {\bibfield  {journal} {\bibinfo  {journal} {New J. Phys}\
  }\textbf {\bibinfo {volume} {21}},\ \bibinfo {pages} {123048} (\bibinfo
  {year} {2019})}\BibitemShut {NoStop}%
\bibitem [{\citenamefont {Lei}\ and\ \citenamefont
  {Ni}(2019)}]{lei2019hydrodynamics}%
  \BibitemOpen
  \bibfield  {author} {\bibinfo {author} {\bibfnamefont {Q.-L.}\ \bibnamefont
  {Lei}}\ and\ \bibinfo {author} {\bibfnamefont {R.}~\bibnamefont {Ni}},\
  }\href@noop {} {\bibfield  {journal} {\bibinfo  {journal} {Proc. Natl. Acad.
  Sci. USA}\ }\textbf {\bibinfo {volume} {116}},\ \bibinfo {pages} {22983}
  (\bibinfo {year} {2019})}\BibitemShut {NoStop}%
\bibitem [{\citenamefont {Lei}\ \emph {et~al.}(2019)\citenamefont {Lei},
  \citenamefont {Ciamarra},\ and\ \citenamefont {Ni}}]{lei2019nonequilibrium}%
  \BibitemOpen
  \bibfield  {author} {\bibinfo {author} {\bibfnamefont {Q.-L.}\ \bibnamefont
  {Lei}}, \bibinfo {author} {\bibfnamefont {M.~P.}\ \bibnamefont {Ciamarra}}, \
  and\ \bibinfo {author} {\bibfnamefont {R.}~\bibnamefont {Ni}},\ }\href@noop
  {} {\bibfield  {journal} {\bibinfo  {journal} {Sci. Adv}\ }\textbf {\bibinfo
  {volume} {5}},\ \bibinfo {pages} {eaau7423} (\bibinfo {year}
  {2019})}\BibitemShut {NoStop}%
\bibitem [{\citenamefont {Chremos}\ and\ \citenamefont
  {Douglas}(2018)}]{chremos2018hidden}%
  \BibitemOpen
  \bibfield  {author} {\bibinfo {author} {\bibfnamefont {A.}~\bibnamefont
  {Chremos}}\ and\ \bibinfo {author} {\bibfnamefont {J.~F.}\ \bibnamefont
  {Douglas}},\ }\href@noop {} {\bibfield  {journal} {\bibinfo  {journal} {Phys.
  Rev. Lett.}\ }\textbf {\bibinfo {volume} {121}},\ \bibinfo {pages} {258002}
  (\bibinfo {year} {2018})}\BibitemShut {NoStop}%
\bibitem [{\citenamefont {Jiao}\ \emph {et~al.}(2014)\citenamefont {Jiao},
  \citenamefont {Lau}, \citenamefont {Hatzikirou}, \citenamefont
  {Meyer-Hermann}, \citenamefont {Corbo},\ and\ \citenamefont
  {Torquato}}]{PhysRevE.89.022721}%
  \BibitemOpen
  \bibfield  {author} {\bibinfo {author} {\bibfnamefont {Y.}~\bibnamefont
  {Jiao}}, \bibinfo {author} {\bibfnamefont {T.}~\bibnamefont {Lau}}, \bibinfo
  {author} {\bibfnamefont {H.}~\bibnamefont {Hatzikirou}}, \bibinfo {author}
  {\bibfnamefont {M.}~\bibnamefont {Meyer-Hermann}}, \bibinfo {author}
  {\bibfnamefont {J.~C.}\ \bibnamefont {Corbo}}, \ and\ \bibinfo {author}
  {\bibfnamefont {S.}~\bibnamefont {Torquato}},\ }\href {\doibase
  10.1103/PhysRevE.89.022721} {\bibfield  {journal} {\bibinfo  {journal} {Phys.
  Rev. E}\ }\textbf {\bibinfo {volume} {89}},\ \bibinfo {pages} {022721}
  (\bibinfo {year} {2014})}\BibitemShut {NoStop}%
\bibitem [{\citenamefont {Mayer}\ \emph {et~al.}(2015)\citenamefont {Mayer},
  \citenamefont {Balasubramanian}, \citenamefont {Mora},\ and\ \citenamefont
  {Walczak}}]{mayer2015well}%
  \BibitemOpen
  \bibfield  {author} {\bibinfo {author} {\bibfnamefont {A.}~\bibnamefont
  {Mayer}}, \bibinfo {author} {\bibfnamefont {V.}~\bibnamefont
  {Balasubramanian}}, \bibinfo {author} {\bibfnamefont {T.}~\bibnamefont
  {Mora}}, \ and\ \bibinfo {author} {\bibfnamefont {A.~M.}\ \bibnamefont
  {Walczak}},\ }\href@noop {} {\bibfield  {journal} {\bibinfo  {journal} {Proc.
  Natl. Acad. Sci. USA}\ }\textbf {\bibinfo {volume} {112}},\ \bibinfo {pages}
  {5950} (\bibinfo {year} {2015})}\BibitemShut {NoStop}%
\bibitem [{\citenamefont {Montgomery}(1973)}]{montgomery1973pair}%
  \BibitemOpen
  \bibfield  {author} {\bibinfo {author} {\bibfnamefont {H.~L.}\ \bibnamefont
  {Montgomery}},\ }in\ \href@noop {} {\emph {\bibinfo {booktitle} {Proc. Symp.
  Pure Math}}},\ Vol.~\bibinfo {volume} {24}\ (\bibinfo {year} {1973})\ pp.\
  \bibinfo {pages} {181--193}\BibitemShut {NoStop}%
\bibitem [{\citenamefont {Torquato}(2018)}]{torquato2018hyperuniform}%
  \BibitemOpen
  \bibfield  {author} {\bibinfo {author} {\bibfnamefont {S.}~\bibnamefont
  {Torquato}},\ }\href@noop {} {\bibfield  {journal} {\bibinfo  {journal}
  {Phys. Rep.}\ } (\bibinfo {year} {2018})}\BibitemShut {NoStop}%
\bibitem [{\citenamefont {Torquato}(2016)}]{PhysRevE.94.022122}%
  \BibitemOpen
  \bibfield  {author} {\bibinfo {author} {\bibfnamefont {S.}~\bibnamefont
  {Torquato}},\ }\href {\doibase 10.1103/PhysRevE.94.022122} {\bibfield
  {journal} {\bibinfo  {journal} {Phys. Rev. E}\ }\textbf {\bibinfo {volume}
  {94}},\ \bibinfo {pages} {022122} (\bibinfo {year} {2016})}\BibitemShut
  {NoStop}%
\bibitem [{\citenamefont {Torquato}(2002)}]{torquato2013random}%
  \BibitemOpen
  \bibfield  {author} {\bibinfo {author} {\bibfnamefont {S.}~\bibnamefont
  {Torquato}},\ }\href@noop {} {\emph {\bibinfo {title} {Random heterogeneous
  materials: microstructure and macroscopic properties}}}\ (\bibinfo
  {publisher} {Springer Science \& Business Media},\ \bibinfo {year}
  {2002})\BibitemShut {NoStop}%
\bibitem [{\citenamefont {Florescu}\ \emph {et~al.}(2009)\citenamefont
  {Florescu}, \citenamefont {Torquato},\ and\ \citenamefont
  {Steinhardt}}]{florescu2009designer}%
  \BibitemOpen
  \bibfield  {author} {\bibinfo {author} {\bibfnamefont {M.}~\bibnamefont
  {Florescu}}, \bibinfo {author} {\bibfnamefont {S.}~\bibnamefont {Torquato}},
  \ and\ \bibinfo {author} {\bibfnamefont {P.~J.}\ \bibnamefont {Steinhardt}},\
  }\href@noop {} {\bibfield  {journal} {\bibinfo  {journal} {Proc. Natl. Acad.
  Sci. USA}\ }\textbf {\bibinfo {volume} {106}},\ \bibinfo {pages} {20658}
  (\bibinfo {year} {2009})}\BibitemShut {NoStop}%
\bibitem [{\citenamefont {Man}\ \emph {et~al.}(2013)\citenamefont {Man},
  \citenamefont {Florescu}, \citenamefont {Williamson}, \citenamefont {He},
  \citenamefont {Hashemizad}, \citenamefont {Leung}, \citenamefont {Liner},
  \citenamefont {Torquato}, \citenamefont {Chaikin},\ and\ \citenamefont
  {Steinhardt}}]{man2013isotropic}%
  \BibitemOpen
  \bibfield  {author} {\bibinfo {author} {\bibfnamefont {W.}~\bibnamefont
  {Man}}, \bibinfo {author} {\bibfnamefont {M.}~\bibnamefont {Florescu}},
  \bibinfo {author} {\bibfnamefont {E.~P.}\ \bibnamefont {Williamson}},
  \bibinfo {author} {\bibfnamefont {Y.}~\bibnamefont {He}}, \bibinfo {author}
  {\bibfnamefont {S.~R.}\ \bibnamefont {Hashemizad}}, \bibinfo {author}
  {\bibfnamefont {B.~Y.}\ \bibnamefont {Leung}}, \bibinfo {author}
  {\bibfnamefont {D.~R.}\ \bibnamefont {Liner}}, \bibinfo {author}
  {\bibfnamefont {S.}~\bibnamefont {Torquato}}, \bibinfo {author}
  {\bibfnamefont {P.~M.}\ \bibnamefont {Chaikin}}, \ and\ \bibinfo {author}
  {\bibfnamefont {P.~J.}\ \bibnamefont {Steinhardt}},\ }\href@noop {}
  {\bibfield  {journal} {\bibinfo  {journal} {Proc. Natl. Acad. Sci. USA}\
  }\textbf {\bibinfo {volume} {110}},\ \bibinfo {pages} {15886} (\bibinfo
  {year} {2013})}\BibitemShut {NoStop}%
\bibitem [{\citenamefont {Haberko}\ \emph {et~al.}(2013)\citenamefont
  {Haberko}, \citenamefont {Muller},\ and\ \citenamefont
  {Scheffold}}]{PhysRevA.88.043822}%
  \BibitemOpen
  \bibfield  {author} {\bibinfo {author} {\bibfnamefont {J.}~\bibnamefont
  {Haberko}}, \bibinfo {author} {\bibfnamefont {N.}~\bibnamefont {Muller}}, \
  and\ \bibinfo {author} {\bibfnamefont {F.}~\bibnamefont {Scheffold}},\ }\href
  {\doibase 10.1103/PhysRevA.88.043822} {\bibfield  {journal} {\bibinfo
  {journal} {Phys. Rev. A}\ }\textbf {\bibinfo {volume} {88}},\ \bibinfo
  {pages} {043822} (\bibinfo {year} {2013})}\BibitemShut {NoStop}%
\bibitem [{\citenamefont {De~Rosa}\ \emph {et~al.}(2015)\citenamefont
  {De~Rosa}, \citenamefont {Auriemma}, \citenamefont {Diletto}, \citenamefont
  {Di~Girolamo}, \citenamefont {Malafronte}, \citenamefont {Morvillo},
  \citenamefont {Zito}, \citenamefont {Rusciano}, \citenamefont {Pesce},\ and\
  \citenamefont {Sasso}}]{C4CP06024E}%
  \BibitemOpen
  \bibfield  {author} {\bibinfo {author} {\bibfnamefont {C.}~\bibnamefont
  {De~Rosa}}, \bibinfo {author} {\bibfnamefont {F.}~\bibnamefont {Auriemma}},
  \bibinfo {author} {\bibfnamefont {C.}~\bibnamefont {Diletto}}, \bibinfo
  {author} {\bibfnamefont {R.}~\bibnamefont {Di~Girolamo}}, \bibinfo {author}
  {\bibfnamefont {A.}~\bibnamefont {Malafronte}}, \bibinfo {author}
  {\bibfnamefont {P.}~\bibnamefont {Morvillo}}, \bibinfo {author}
  {\bibfnamefont {G.}~\bibnamefont {Zito}}, \bibinfo {author} {\bibfnamefont
  {G.}~\bibnamefont {Rusciano}}, \bibinfo {author} {\bibfnamefont
  {G.}~\bibnamefont {Pesce}}, \ and\ \bibinfo {author} {\bibfnamefont
  {A.}~\bibnamefont {Sasso}},\ }\href {\doibase 10.1039/C4CP06024E} {\bibfield
  {journal} {\bibinfo  {journal} {Phys. Chem. Chem. Phys.}\ }\textbf {\bibinfo
  {volume} {17}},\ \bibinfo {pages} {8061} (\bibinfo {year}
  {2015})}\BibitemShut {NoStop}%
\bibitem [{\citenamefont {Piechulla}\ \emph {et~al.}(2018)\citenamefont
  {Piechulla}, \citenamefont {Muehlenbein}, \citenamefont {Wehrspohn},
  \citenamefont {Nanz}, \citenamefont {Abass}, \citenamefont {Rockstuhl},\ and\
  \citenamefont {Sprafke}}]{piechulla2018fabrication}%
  \BibitemOpen
  \bibfield  {author} {\bibinfo {author} {\bibfnamefont {P.~M.}\ \bibnamefont
  {Piechulla}}, \bibinfo {author} {\bibfnamefont {L.}~\bibnamefont
  {Muehlenbein}}, \bibinfo {author} {\bibfnamefont {R.~B.}\ \bibnamefont
  {Wehrspohn}}, \bibinfo {author} {\bibfnamefont {S.}~\bibnamefont {Nanz}},
  \bibinfo {author} {\bibfnamefont {A.}~\bibnamefont {Abass}}, \bibinfo
  {author} {\bibfnamefont {C.}~\bibnamefont {Rockstuhl}}, \ and\ \bibinfo
  {author} {\bibfnamefont {A.}~\bibnamefont {Sprafke}},\ }\href@noop {}
  {\bibfield  {journal} {\bibinfo  {journal} {Adv. Opt. Mater}\ }\textbf
  {\bibinfo {volume} {6}},\ \bibinfo {pages} {1701272} (\bibinfo {year}
  {2018})}\BibitemShut {NoStop}%
\bibitem [{\citenamefont {Bigourdan}\ \emph {et~al.}(2019)\citenamefont
  {Bigourdan}, \citenamefont {Pierrat},\ and\ \citenamefont
  {Carminati}}]{bigourdan2019enhanced}%
  \BibitemOpen
  \bibfield  {author} {\bibinfo {author} {\bibfnamefont {F.}~\bibnamefont
  {Bigourdan}}, \bibinfo {author} {\bibfnamefont {R.}~\bibnamefont {Pierrat}},
  \ and\ \bibinfo {author} {\bibfnamefont {R.}~\bibnamefont {Carminati}},\
  }\href@noop {} {\bibfield  {journal} {\bibinfo  {journal} {Opt. Express}\
  }\textbf {\bibinfo {volume} {27}},\ \bibinfo {pages} {8666} (\bibinfo {year}
  {2019})}\BibitemShut {NoStop}%
\bibitem [{\citenamefont {Gorsky}\ \emph {et~al.}(2019)\citenamefont {Gorsky},
  \citenamefont {Britton}, \citenamefont {Chen}, \citenamefont {Montaner},
  \citenamefont {Lenef}, \citenamefont {Raukas},\ and\ \citenamefont
  {Dal~Negro}}]{gorsky2019engineered}%
  \BibitemOpen
  \bibfield  {author} {\bibinfo {author} {\bibfnamefont {S.}~\bibnamefont
  {Gorsky}}, \bibinfo {author} {\bibfnamefont {W.}~\bibnamefont {Britton}},
  \bibinfo {author} {\bibfnamefont {Y.}~\bibnamefont {Chen}}, \bibinfo {author}
  {\bibfnamefont {J.}~\bibnamefont {Montaner}}, \bibinfo {author}
  {\bibfnamefont {A.}~\bibnamefont {Lenef}}, \bibinfo {author} {\bibfnamefont
  {M.}~\bibnamefont {Raukas}}, \ and\ \bibinfo {author} {\bibfnamefont
  {L.}~\bibnamefont {Dal~Negro}},\ }\href@noop {} {\bibfield  {journal}
  {\bibinfo  {journal} {APL Photonics}\ }\textbf {\bibinfo {volume} {4}},\
  \bibinfo {pages} {110801} (\bibinfo {year} {2019})}\BibitemShut {NoStop}%
\bibitem [{\citenamefont {Romero-Garc{\'\i}a}\ \emph
  {et~al.}(2019)\citenamefont {Romero-Garc{\'\i}a}, \citenamefont {Lamothe},
  \citenamefont {Theocharis}, \citenamefont {Richoux},\ and\ \citenamefont
  {Garc{\'\i}a-Raffi}}]{romero2019stealth}%
  \BibitemOpen
  \bibfield  {author} {\bibinfo {author} {\bibfnamefont {V.}~\bibnamefont
  {Romero-Garc{\'\i}a}}, \bibinfo {author} {\bibfnamefont {N.}~\bibnamefont
  {Lamothe}}, \bibinfo {author} {\bibfnamefont {G.}~\bibnamefont {Theocharis}},
  \bibinfo {author} {\bibfnamefont {O.}~\bibnamefont {Richoux}}, \ and\
  \bibinfo {author} {\bibfnamefont {L.}~\bibnamefont {Garc{\'\i}a-Raffi}},\
  }\href@noop {} {\bibfield  {journal} {\bibinfo  {journal} {Phys. Rev. Appl}\
  }\textbf {\bibinfo {volume} {11}},\ \bibinfo {pages} {054076} (\bibinfo
  {year} {2019})}\BibitemShut {NoStop}%
\bibitem [{\citenamefont {Noh}\ \emph {et~al.}(2010)\citenamefont {Noh},
  \citenamefont {Liew}, \citenamefont {Saranathan}, \citenamefont {Mochrie},
  \citenamefont {Prum}, \citenamefont {Dufresne},\ and\ \citenamefont
  {Cao}}]{noh2010noniridescent}%
  \BibitemOpen
  \bibfield  {author} {\bibinfo {author} {\bibfnamefont {H.}~\bibnamefont
  {Noh}}, \bibinfo {author} {\bibfnamefont {S.~F.}\ \bibnamefont {Liew}},
  \bibinfo {author} {\bibfnamefont {V.}~\bibnamefont {Saranathan}}, \bibinfo
  {author} {\bibfnamefont {S.~G.}\ \bibnamefont {Mochrie}}, \bibinfo {author}
  {\bibfnamefont {R.~O.}\ \bibnamefont {Prum}}, \bibinfo {author}
  {\bibfnamefont {E.~R.}\ \bibnamefont {Dufresne}}, \ and\ \bibinfo {author}
  {\bibfnamefont {H.}~\bibnamefont {Cao}},\ }\href@noop {} {\bibfield
  {journal} {\bibinfo  {journal} {Adv. Mater}\ }\textbf {\bibinfo {volume}
  {22}},\ \bibinfo {pages} {2871} (\bibinfo {year} {2010})}\BibitemShut
  {NoStop}%
\bibitem [{\citenamefont {Chung}\ \emph {et~al.}(2012)\citenamefont {Chung},
  \citenamefont {Yu}, \citenamefont {Heo}, \citenamefont {Shim}, \citenamefont
  {Yang}, \citenamefont {Han}, \citenamefont {Lee}, \citenamefont {Jin},
  \citenamefont {Lee}, \citenamefont {Park},\ and\ \citenamefont
  {Shin}}]{chung2012flexible}%
  \BibitemOpen
  \bibfield  {author} {\bibinfo {author} {\bibfnamefont {K.}~\bibnamefont
  {Chung}}, \bibinfo {author} {\bibfnamefont {S.}~\bibnamefont {Yu}}, \bibinfo
  {author} {\bibfnamefont {C.-J.}\ \bibnamefont {Heo}}, \bibinfo {author}
  {\bibfnamefont {J.~W.}\ \bibnamefont {Shim}}, \bibinfo {author}
  {\bibfnamefont {S.-M.}\ \bibnamefont {Yang}}, \bibinfo {author}
  {\bibfnamefont {M.~G.}\ \bibnamefont {Han}}, \bibinfo {author} {\bibfnamefont
  {H.-S.}\ \bibnamefont {Lee}}, \bibinfo {author} {\bibfnamefont
  {Y.}~\bibnamefont {Jin}}, \bibinfo {author} {\bibfnamefont {S.~Y.}\
  \bibnamefont {Lee}}, \bibinfo {author} {\bibfnamefont {N.}~\bibnamefont
  {Park}}, \ and\ \bibinfo {author} {\bibfnamefont {J.~H.}\ \bibnamefont
  {Shin}},\ }\href@noop {} {\bibfield  {journal} {\bibinfo  {journal} {Adv.
  Mater}\ }\textbf {\bibinfo {volume} {24}},\ \bibinfo {pages} {2375} (\bibinfo
  {year} {2012})}\BibitemShut {NoStop}%
\bibitem [{\citenamefont {Ricouvier}\ \emph {et~al.}(2017)\citenamefont
  {Ricouvier}, \citenamefont {Pierrat}, \citenamefont {Carminati},
  \citenamefont {Tabeling},\ and\ \citenamefont
  {Yazhgur}}]{ricouvier2017optimizing}%
  \BibitemOpen
  \bibfield  {author} {\bibinfo {author} {\bibfnamefont {J.}~\bibnamefont
  {Ricouvier}}, \bibinfo {author} {\bibfnamefont {R.}~\bibnamefont {Pierrat}},
  \bibinfo {author} {\bibfnamefont {R.}~\bibnamefont {Carminati}}, \bibinfo
  {author} {\bibfnamefont {P.}~\bibnamefont {Tabeling}}, \ and\ \bibinfo
  {author} {\bibfnamefont {P.}~\bibnamefont {Yazhgur}},\ }\href@noop {}
  {\bibfield  {journal} {\bibinfo  {journal} {Phys. Rev. Lett.}\ }\textbf
  {\bibinfo {volume} {119}},\ \bibinfo {pages} {208001} (\bibinfo {year}
  {2017})}\BibitemShut {NoStop}%
\bibitem [{\citenamefont {Berthier}\ \emph {et~al.}(2011)\citenamefont
  {Berthier}, \citenamefont {Chaudhuri}, \citenamefont {Coulais}, \citenamefont
  {Dauchot},\ and\ \citenamefont {Sollich}}]{berthier2011suppressed}%
  \BibitemOpen
  \bibfield  {author} {\bibinfo {author} {\bibfnamefont {L.}~\bibnamefont
  {Berthier}}, \bibinfo {author} {\bibfnamefont {P.}~\bibnamefont {Chaudhuri}},
  \bibinfo {author} {\bibfnamefont {C.}~\bibnamefont {Coulais}}, \bibinfo
  {author} {\bibfnamefont {O.}~\bibnamefont {Dauchot}}, \ and\ \bibinfo
  {author} {\bibfnamefont {P.}~\bibnamefont {Sollich}},\ }\href@noop {}
  {\bibfield  {journal} {\bibinfo  {journal} {Phys. Rev. Lett.}\ }\textbf
  {\bibinfo {volume} {106}},\ \bibinfo {pages} {120601} (\bibinfo {year}
  {2011})}\BibitemShut {NoStop}%
\bibitem [{\citenamefont {Weijs}\ and\ \citenamefont
  {Bartolo}(2017)}]{weijs2017mixing}%
  \BibitemOpen
  \bibfield  {author} {\bibinfo {author} {\bibfnamefont {J.~H.}\ \bibnamefont
  {Weijs}}\ and\ \bibinfo {author} {\bibfnamefont {D.}~\bibnamefont
  {Bartolo}},\ }\href@noop {} {\bibfield  {journal} {\bibinfo  {journal} {Phys.
  Rev. Lett.}\ }\textbf {\bibinfo {volume} {119}},\ \bibinfo {pages} {048002}
  (\bibinfo {year} {2017})}\BibitemShut {NoStop}%
\bibitem [{\citenamefont {Wilken}\ \emph {et~al.}(2020)\citenamefont {Wilken},
  \citenamefont {Guerra}, \citenamefont {Pine},\ and\ \citenamefont
  {Chaikin}}]{wilken2020hyperuniform}%
  \BibitemOpen
  \bibfield  {author} {\bibinfo {author} {\bibfnamefont {S.}~\bibnamefont
  {Wilken}}, \bibinfo {author} {\bibfnamefont {R.~E.}\ \bibnamefont {Guerra}},
  \bibinfo {author} {\bibfnamefont {D.~J.}\ \bibnamefont {Pine}}, \ and\
  \bibinfo {author} {\bibfnamefont {P.~M.}\ \bibnamefont {Chaikin}},\
  }\href@noop {} {\bibfield  {journal} {\bibinfo  {journal} {arXiv preprint
  arXiv:2002.04499}\ } (\bibinfo {year} {2020})}\BibitemShut {NoStop}%
\bibitem [{\citenamefont {Ma}\ and\ \citenamefont
  {Torquato}(2017)}]{ma2017random}%
  \BibitemOpen
  \bibfield  {author} {\bibinfo {author} {\bibfnamefont {Z.}~\bibnamefont
  {Ma}}\ and\ \bibinfo {author} {\bibfnamefont {S.}~\bibnamefont {Torquato}},\
  }\href@noop {} {\bibfield  {journal} {\bibinfo  {journal} {J. Appl. Phys}\
  }\textbf {\bibinfo {volume} {121}},\ \bibinfo {pages} {244904} (\bibinfo
  {year} {2017})}\BibitemShut {NoStop}%
\bibitem [{\citenamefont {Salvalaglio}\ \emph {et~al.}(2019)\citenamefont
  {Salvalaglio}, \citenamefont {Bouabdellaoui}, \citenamefont {Bollani},
  \citenamefont {Benali}, \citenamefont {Favre}, \citenamefont {Claude},
  \citenamefont {Wenger}, \citenamefont {de~Anna}, \citenamefont {Intonti},
  \citenamefont {Voigt},\ and\ \citenamefont
  {Abbarchi}}]{salvalaglio2019hyperuniform}%
  \BibitemOpen
  \bibfield  {author} {\bibinfo {author} {\bibfnamefont {M.}~\bibnamefont
  {Salvalaglio}}, \bibinfo {author} {\bibfnamefont {M.}~\bibnamefont
  {Bouabdellaoui}}, \bibinfo {author} {\bibfnamefont {M.}~\bibnamefont
  {Bollani}}, \bibinfo {author} {\bibfnamefont {A.}~\bibnamefont {Benali}},
  \bibinfo {author} {\bibfnamefont {L.}~\bibnamefont {Favre}}, \bibinfo
  {author} {\bibfnamefont {J.-B.}\ \bibnamefont {Claude}}, \bibinfo {author}
  {\bibfnamefont {J.}~\bibnamefont {Wenger}}, \bibinfo {author} {\bibfnamefont
  {P.}~\bibnamefont {de~Anna}}, \bibinfo {author} {\bibfnamefont
  {F.}~\bibnamefont {Intonti}}, \bibinfo {author} {\bibfnamefont
  {A.}~\bibnamefont {Voigt}}, \ and\ \bibinfo {author} {\bibfnamefont
  {M.}~\bibnamefont {Abbarchi}},\ }\href@noop {} {\bibfield  {journal}
  {\bibinfo  {journal} {arXiv preprint arXiv:1912.02952}\ } (\bibinfo {year}
  {2019})}\BibitemShut {NoStop}%
\bibitem [{\citenamefont {Atkinson}\ \emph {et~al.}(2016)\citenamefont
  {Atkinson}, \citenamefont {Zhang}, \citenamefont {Hopkins},\ and\
  \citenamefont {Torquato}}]{atkinson2016critical}%
  \BibitemOpen
  \bibfield  {author} {\bibinfo {author} {\bibfnamefont {S.}~\bibnamefont
  {Atkinson}}, \bibinfo {author} {\bibfnamefont {G.}~\bibnamefont {Zhang}},
  \bibinfo {author} {\bibfnamefont {A.~B.}\ \bibnamefont {Hopkins}}, \ and\
  \bibinfo {author} {\bibfnamefont {S.}~\bibnamefont {Torquato}},\ }\href@noop
  {} {\bibfield  {journal} {\bibinfo  {journal} {Phys. Rev. E}\ }\textbf
  {\bibinfo {volume} {94}},\ \bibinfo {pages} {012902} (\bibinfo {year}
  {2016})}\BibitemShut {NoStop}%
\bibitem [{\citenamefont {Chen}\ \emph {et~al.}(2018)\citenamefont {Chen},
  \citenamefont {Lomba},\ and\ \citenamefont {Torquato}}]{chen2018binary}%
  \BibitemOpen
  \bibfield  {author} {\bibinfo {author} {\bibfnamefont {D.}~\bibnamefont
  {Chen}}, \bibinfo {author} {\bibfnamefont {E.}~\bibnamefont {Lomba}}, \ and\
  \bibinfo {author} {\bibfnamefont {S.}~\bibnamefont {Torquato}},\ }\href@noop
  {} {\bibfield  {journal} {\bibinfo  {journal} {Phys. Chem. Chem. Phys.}\
  }\textbf {\bibinfo {volume} {20}},\ \bibinfo {pages} {17557} (\bibinfo {year}
  {2018})}\BibitemShut {NoStop}%
\bibitem [{\citenamefont {Ebert}\ \emph {et~al.}(2009)\citenamefont {Ebert},
  \citenamefont {Maret},\ and\ \citenamefont {Keim}}]{ebert2009partial}%
  \BibitemOpen
  \bibfield  {author} {\bibinfo {author} {\bibfnamefont {F.}~\bibnamefont
  {Ebert}}, \bibinfo {author} {\bibfnamefont {G.}~\bibnamefont {Maret}}, \ and\
  \bibinfo {author} {\bibfnamefont {P.}~\bibnamefont {Keim}},\ }\href@noop {}
  {\bibfield  {journal} {\bibinfo  {journal} {Eur. Phys. J. E}\ }\textbf
  {\bibinfo {volume} {29}},\ \bibinfo {pages} {311} (\bibinfo {year}
  {2009})}\BibitemShut {NoStop}%
\bibitem [{\citenamefont {Lin}\ \emph {et~al.}(2006)\citenamefont {Lin},
  \citenamefont {Zheng},\ and\ \citenamefont {Trimper}}]{lin2006computer}%
  \BibitemOpen
  \bibfield  {author} {\bibinfo {author} {\bibfnamefont {S.}~\bibnamefont
  {Lin}}, \bibinfo {author} {\bibfnamefont {B.}~\bibnamefont {Zheng}}, \ and\
  \bibinfo {author} {\bibfnamefont {S.}~\bibnamefont {Trimper}},\ }\href@noop
  {} {\bibfield  {journal} {\bibinfo  {journal} {Phys. Rev. E}\ }\textbf
  {\bibinfo {volume} {73}},\ \bibinfo {pages} {066106} (\bibinfo {year}
  {2006})}\BibitemShut {NoStop}%
\bibitem [{\citenamefont {Kapfer}\ and\ \citenamefont
  {Krauth}(2015)}]{kapfer2015two}%
  \BibitemOpen
  \bibfield  {author} {\bibinfo {author} {\bibfnamefont {S.~C.}\ \bibnamefont
  {Kapfer}}\ and\ \bibinfo {author} {\bibfnamefont {W.}~\bibnamefont
  {Krauth}},\ }\href@noop {} {\bibfield  {journal} {\bibinfo  {journal} {Phys.
  Rev. Lett.}\ }\textbf {\bibinfo {volume} {114}},\ \bibinfo {pages} {035702}
  (\bibinfo {year} {2015})}\BibitemShut {NoStop}%
\bibitem [{\citenamefont {Zahn}\ \emph {et~al.}(1999)\citenamefont {Zahn},
  \citenamefont {Lenke},\ and\ \citenamefont {Maret}}]{zahn1999two}%
  \BibitemOpen
  \bibfield  {author} {\bibinfo {author} {\bibfnamefont {K.}~\bibnamefont
  {Zahn}}, \bibinfo {author} {\bibfnamefont {R.}~\bibnamefont {Lenke}}, \ and\
  \bibinfo {author} {\bibfnamefont {G.}~\bibnamefont {Maret}},\ }\href@noop {}
  {\bibfield  {journal} {\bibinfo  {journal} {Phys. Rev. Lett.}\ }\textbf
  {\bibinfo {volume} {82}},\ \bibinfo {pages} {2721} (\bibinfo {year}
  {1999})}\BibitemShut {NoStop}%
\bibitem [{\citenamefont {Kelleher}\ \emph {et~al.}(2017)\citenamefont
  {Kelleher}, \citenamefont {Guerra}, \citenamefont {Hollingsworth},\ and\
  \citenamefont {Chaikin}}]{kelleher2017phase}%
  \BibitemOpen
  \bibfield  {author} {\bibinfo {author} {\bibfnamefont {C.~P.}\ \bibnamefont
  {Kelleher}}, \bibinfo {author} {\bibfnamefont {R.~E.}\ \bibnamefont
  {Guerra}}, \bibinfo {author} {\bibfnamefont {A.~D.}\ \bibnamefont
  {Hollingsworth}}, \ and\ \bibinfo {author} {\bibfnamefont {P.~M.}\
  \bibnamefont {Chaikin}},\ }\href@noop {} {\bibfield  {journal} {\bibinfo
  {journal} {Phys. Rev. E}\ }\textbf {\bibinfo {volume} {95}},\ \bibinfo
  {pages} {022602} (\bibinfo {year} {2017})}\BibitemShut {NoStop}%
\bibitem [{\citenamefont {Halperin}\ and\ \citenamefont
  {Nelson}(1978)}]{halperin1978theory}%
  \BibitemOpen
  \bibfield  {author} {\bibinfo {author} {\bibfnamefont {B.}~\bibnamefont
  {Halperin}}\ and\ \bibinfo {author} {\bibfnamefont {D.~R.}\ \bibnamefont
  {Nelson}},\ }\href@noop {} {\bibfield  {journal} {\bibinfo  {journal} {Phys.
  Rev. Lett.}\ }\textbf {\bibinfo {volume} {41}},\ \bibinfo {pages} {121}
  (\bibinfo {year} {1978})}\BibitemShut {NoStop}%
\bibitem [{\citenamefont {Hoffmann}\ \emph
  {et~al.}(2006{\natexlab{a}})\citenamefont {Hoffmann}, \citenamefont {Likos},\
  and\ \citenamefont {L\"{o}wen}}]{Hoffman2006}%
  \BibitemOpen
  \bibfield  {author} {\bibinfo {author} {\bibfnamefont {N.}~\bibnamefont
  {Hoffmann}}, \bibinfo {author} {\bibfnamefont {C.~N.}\ \bibnamefont {Likos}},
  \ and\ \bibinfo {author} {\bibfnamefont {H.}~\bibnamefont {L\"{o}wen}},\
  }\href {\doibase 10.1088/0953-8984/18/45/007} {\bibfield  {journal} {\bibinfo
   {journal} {J Phys : Condens Matter}\ }\textbf {\bibinfo {volume} {18}},\
  \bibinfo {pages} {10193} (\bibinfo {year} {2006}{\natexlab{a}})}\BibitemShut
  {NoStop}%
\bibitem [{\citenamefont {Hoffmann}\ \emph
  {et~al.}(2006{\natexlab{b}})\citenamefont {Hoffmann}, \citenamefont {Ebert},
  \citenamefont {Likos}, \citenamefont {L{\"o}wen},\ and\ \citenamefont
  {Maret}}]{hoffmann2006partial}%
  \BibitemOpen
  \bibfield  {author} {\bibinfo {author} {\bibfnamefont {N.}~\bibnamefont
  {Hoffmann}}, \bibinfo {author} {\bibfnamefont {F.}~\bibnamefont {Ebert}},
  \bibinfo {author} {\bibfnamefont {C.~N.}\ \bibnamefont {Likos}}, \bibinfo
  {author} {\bibfnamefont {H.}~\bibnamefont {L{\"o}wen}}, \ and\ \bibinfo
  {author} {\bibfnamefont {G.}~\bibnamefont {Maret}},\ }\href@noop {}
  {\bibfield  {journal} {\bibinfo  {journal} {Phys. Rev. Lett.}\ }\textbf
  {\bibinfo {volume} {97}},\ \bibinfo {pages} {078301} (\bibinfo {year}
  {2006}{\natexlab{b}})}\BibitemShut {NoStop}%
\bibitem [{\citenamefont {Grigera}\ and\ \citenamefont
  {Parisi}(2001)}]{grigera2001fast}%
  \BibitemOpen
  \bibfield  {author} {\bibinfo {author} {\bibfnamefont {T.~S.}\ \bibnamefont
  {Grigera}}\ and\ \bibinfo {author} {\bibfnamefont {G.}~\bibnamefont
  {Parisi}},\ }\href@noop {} {\bibfield  {journal} {\bibinfo  {journal} {Phys.
  Rev. E}\ }\textbf {\bibinfo {volume} {63}},\ \bibinfo {pages} {045102}
  (\bibinfo {year} {2001})}\BibitemShut {NoStop}%
\bibitem [{\citenamefont {Rogers}\ and\ \citenamefont
  {Young}(1984)}]{Rogers1984}%
  \BibitemOpen
  \bibfield  {author} {\bibinfo {author} {\bibfnamefont {F.~J.}\ \bibnamefont
  {Rogers}}\ and\ \bibinfo {author} {\bibfnamefont {D.~A.}\ \bibnamefont
  {Young}},\ }\href {\doibase 10.1103/PhysRevA.30.999} {\bibfield  {journal}
  {\bibinfo  {journal} {Phys. Rev. A}\ }\textbf {\bibinfo {volume} {30}},\
  \bibinfo {pages} {999 } (\bibinfo {year} {1984})}\BibitemShut {NoStop}%
\bibitem [{\citenamefont {Lomba}\ \emph {et~al.}(2017)\citenamefont {Lomba},
  \citenamefont {Weis},\ and\ \citenamefont {Torquato}}]{Lomba2017}%
  \BibitemOpen
  \bibfield  {author} {\bibinfo {author} {\bibfnamefont {E.}~\bibnamefont
  {Lomba}}, \bibinfo {author} {\bibfnamefont {J.~J.}\ \bibnamefont {Weis}}, \
  and\ \bibinfo {author} {\bibfnamefont {S.}~\bibnamefont {Torquato}},\ }\href
  {\doibase 10.1103/PhysRevE.96.062126} {\bibfield  {journal} {\bibinfo
  {journal} {Phys. Rev. E}\ }\textbf {\bibinfo {volume} {96}},\ \bibinfo
  {pages} {062126} (\bibinfo {year} {2017})}\BibitemShut {NoStop}%
\bibitem [{sup()}]{supplement}%
  \BibitemOpen
  \href@noop {} {}\bibinfo {note} {See Supplemental Material at [URL] for
  additional information.}\BibitemShut {Stop}%
\bibitem [{\citenamefont {Ashcroft}\ and\ \citenamefont
  {Langreth}(1967)}]{ashcroft1967structure}%
  \BibitemOpen
  \bibfield  {author} {\bibinfo {author} {\bibfnamefont {N.}~\bibnamefont
  {Ashcroft}}\ and\ \bibinfo {author} {\bibfnamefont {D.~C.}\ \bibnamefont
  {Langreth}},\ }\href@noop {} {\bibfield  {journal} {\bibinfo  {journal}
  {Phys. Rev}\ }\textbf {\bibinfo {volume} {156}},\ \bibinfo {pages} {685}
  (\bibinfo {year} {1967})}\BibitemShut {NoStop}%
\bibitem [{\citenamefont {Khrapak}\ \emph {et~al.}(2018)\citenamefont
  {Khrapak}, \citenamefont {Kryuchkov},\ and\ \citenamefont
  {Yurchenko}}]{khrapak2018thermodynamics}%
  \BibitemOpen
  \bibfield  {author} {\bibinfo {author} {\bibfnamefont {S.~A.}\ \bibnamefont
  {Khrapak}}, \bibinfo {author} {\bibfnamefont {N.~P.}\ \bibnamefont
  {Kryuchkov}}, \ and\ \bibinfo {author} {\bibfnamefont {S.~O.}\ \bibnamefont
  {Yurchenko}},\ }\href@noop {} {\bibfield  {journal} {\bibinfo  {journal}
  {Phys. Rev. E}\ }\textbf {\bibinfo {volume} {97}},\ \bibinfo {pages} {022616}
  (\bibinfo {year} {2018})}\BibitemShut {NoStop}%
\end{thebibliography}
\end{document}